# Fluorine-substitution-dependent phase diagram and superconducting properties of Sm-based oxypnictides synthesized by a high-pressure growth technique

Mohammad Azam[1], Tatiana Zajarniuk[2], Ryszard Diduszko[2,3] Taras Palasyuk[4], Cezariusz Jastrzebski[4], Andrzej Szewczyk[2], Hiraku Ogino[5], Shiv J. Singh[1*]

*[1]Institute of High Pressure Physics (IHPP), Polish Academy of Sciences, Sokołowska 29/37, 01-142 Warsaw, Poland*

*[2]Institute of Physics, Polish Academy of Sciences, Aleja Lotników 32/46, 02-668 Warsaw, Poland*

*[3]National Centre for Nuclear Research, ul. Andrzeja Sołtana 7/3, 05-400 Otwock-Swierk, Poland*

*[4]Faculty of Physics, Warsaw University of Technology, Koszykowa 75, 00-662 Warsaw, Poland*

*[5]Core Electronics Technology Research Institute, National Institute of Advanced Industrial Science and Technology (AIST), Tsukuba, Ibaraki 305-8568, Japan*

***Corresponding author:**
Email: sjs@unipress.waw.pl

https://orcid.org/0000-0001-5769-1787

# Abstract

A series of $SmFeAsO_{1-x}F_x$ (Sm1111) bulk samples ($0.05 \leq x \leq 0.40$) are synthesized by an *in-situ* cubic-anvil high-pressure (CA-HP) technique at 4 GPa and systematically characterized through structural, microstructural, Raman, transport, and magnetic measurements. Structural analysis confirms that the tetragonal Sm1111 phase remains dominant across the entire substitution range, with lattice parameters decreasing smoothly as fluorine content increases, demonstrating effective incorporation of F even in the overdoped regime ($x = 0.4$). Raman spectroscopy provides complementary, local-phase-sensitive evidence that supports the structural analysis and confirms fluorine substitution in the main Sm1111 phase. In the underdoped region ($0.05 \leq x < 0.2$), the superconducting transition temperature ($T_c$) is enhanced by 10-17 K and the critical current density ($J_c$) is increased by up to an order of magnitude compared with conventionally synthesized samples at the ambient pressure. The upper critical field ($H_{c2}$), estimated using the Werthamer–Helfand–Hohenberg model, reaches values approaching 200 T, reflecting strong paramagnetic limitation and multiband effects. Thermally activated flux-flow analysis reveals a power-law field dependence of the activation energy, consistent with collective vortex pinning in polycrystalline iron-based superconductors. The superconducting phase diagram constructed from $T_c$ and $J_c$ versus fluorine content reveals a dome-like trend, with a maximum $T_c$ of 57 K and $J_c$ of $10^4$ A cm$^{-2}$ at the optimal doped region. Comparison with conventionally synthesized Sm1111 samples demonstrates that high-pressure growth significantly extends the effective fluorine substitution range and enhances the superconducting performance, particularly in the underdoped regime. These results establish high-pressure fluorine substitution as an effective materials-engineering approach for tuning the phase diagram and optimizing the superconducting properties of Sm-based oxypnictide.

## Introduction:

The discovery of iron-based superconductors (FBS), particularly the layered oxypnictide *RE*FeAsO (*RE* = rare earth; 1111) systems [1], has significantly advanced the field of high-temperature superconductivity [2, 3]. Among these, fluorine-doped $SmFeAsO_{1-x}F_x$ (Sm1111) exhibits the highest superconducting transition temperature ($T_c \approx$ 57-58 K) reported for FBS [4, 5, 6, 7], along with exceptionally high upper critical fields ($\mu_0 H_{c2} \approx$ 200-300 T) [5, 8]. These properties make Sm1111 a promising candidate for high-magnetic-field applications without the need for liquid helium cooling [5, 9, 10]. Despite the few studies based on the growth of F-doped Sm1111 single crystals with critical current densities ($J_c$) of ~$10^6$ A/cm² at low temperatures [11], synthesizing high-quality bulk and single-crystalline samples remains challenging [9]. Structurally, the 1111-type compounds comprise electronically active FeAs layers, which host superconductivity, alternated with *RE*O charge-reservoir layers [12]. The parent *RE*FeAsO compound is non-superconducting and undergoes an antiferromagnetic spin-density-wave (SDW) transition near 150 K [4]. Superconductivity emerges upon carrier doping that suppresses this SDW order [13]. In SmFeAsO, electron doping via partial substitution of $O^{2-}$ by $F^-$ in the SmO layer [14] transfers charge carriers into the Fe 3d bands, forming quasi-two-dimensional Fermi surfaces consisting of hole pockets at the Γ point and electron pockets at the M point [12]. Nesting between these pockets enhances interband scattering and promotes spin fluctuations, consistent with an s± pairing symmetry mediated by spin fluctuations [15, 16] [17, 18]. Fluorine substitution also modifies the local FeAs-layers geometry, tuning geometric parameters such as the As-Fe-As bond angle (ideal ≈109.5°) and the pnictogen height ($h_{As}$) above the Fe plane, which correlate strongly with superconducting pairing strength [13] [19] [6]. Approaching the ideal tetrahedral geometry and optimizing $h_{As}$ are believed to enhance the density of states and spin-fluctuation spectrum, favoring higher $T_c$ [20] [21]. Achieving this structural optimization experimentally, however, is highly nontrivial.

Despite these advances, conventional synthesis process at ambient pressure (CSP) limits fluorine solubility due to volatility, competing secondary phases (SmOF, $Sm_2O_3$, SmAs, FeAs), and low phase density, restricting reliable F substitution to $x \lesssim$ 0.20-0.25 in Sm1111 [3] [9]. This limitation prevents exploration of the overdoped regime ($x$ > 0.25), leaving important regions of the phase diagram largely unstudied. Similar limitations are observed in La1111, where F substitution is restricted to $x \approx 0.2$, whereas hydrogen substitution ($LaFeAsO_{1-x}H_x$) enables the substitution range up to $x \approx 0.4$ and reveals a double-dome superconducting phase diagram [22, 23]. However, because $H^-$ has a much smaller ionic radius than $O^{2-}$, H substitution introduces strong chemical-pressure effects, making it challenging to disentangle the contributions of electron doping from structural changes [22, 24, 23]. In contrast, $F^-$ has an ionic radius closer to $O^{2-}$, so high-pressure F substitution provides primarily electron doping with minimal lattice perturbation, offering a cleaner route to study intrinsic superconducting evolution. Notably, in Sm1111, double-dome superconductivity has also been observed in H-substituted systems where As is partially replaced by P [25], further highlighting the significance of exploring the overdoped region. On the other hand, low-temperature CSP has improved fluorine retention and achieved $T_c$ values close to 58 K, suggesting that higher F content can be stabilized when volatility and parasitic reactions are minimized. Nevertheless, the corresponding $J_c$ remains on the order of $10^3$ A/cm² at 5 K [5]. In contrast, high-temperature CSP (~1200°C) improves $J_c$ by about one order of magnitude but reduces $T_c$ by ~4 K due to enhanced fluorine evaporation [26]. Thus, achieving both high $T_c$ and high $J_c$ simultaneously via CSP remains difficult.

To overcome these limitations, high-pressure (HP) synthesis has emerged as a powerful strategy for improving sample density, phase purity, and dopant solubility while suppressing secondary phases [27] [28, 29]. Using HP techniques, Ren *et al.* synthesized $SmFeAsO_{0.90}F_{0.10}$ with $T_c \approx 55$ K [30] and $NdFeAsO_{0.89}F_{0.10}$ [31] with $T_c = 51.9$ K under 6 GPa. In addition, oxygen-deficient $REFeAsO_{1-x}$ phases can be stabilized only under high pressure [32]. More recently, HP-induced enhancements of superconductivity have also been reported for $FeSe_{0.5}Te_{0.5}$ [28], $CaKFe_4As_4$ [33], and F-doped SmFeAsO [34, 35]. Spark plasma sintering (SPS) has also yielded denser and cleaner Sm1111 samples near optimal substitution ($x \approx 0.20$) [36]. Despite these advances, most HP and SPS studies remain confined to compositions near optimal substitution (~20%), leaving the underdoped and, especially, overdoped regimes ($x > 0.25$) largely unexplored, primarily due to the volatility of fluorine under ambient conditions. A systematic investigation across both underdoped and overdoped regimes under identical high-pressure conditions is therefore crucible for establishing the complete phase diagram and understand the interplay between crystal structure, electron doping, and superconducting properties.

In this work, we present a comprehensive investigation of polycrystalline $SmFeAsO_{1-x}F_x$ ($0.05 \leq x \leq 0.40$) synthesized by a cubic-anvil high-pressure (CA-HP) technique at 4 GPa. By maintaining identical synthesis conditions across the entire composition range, we systematically evaluate the effects of fluorine substitution on crystal structure, phase stability, and superconducting properties, including the transition temperature, upper critical field, critical current density, vortex-activation energy, and phase composition. Comparison with conventionally synthesized samples demonstrates that high-pressure processing substantially extends the accessible fluorine-doping range into the overdoped regime, yielding dense and homogeneous Sm-1111 bulks with enhanced superconducting performance and providing new insight into the evolution of the electron-doped phase diagram.

## Experimental details:

$SmFeAsO_{1-x}F_x$ bulks were prepared using a single-step *in-situ* method under cubic-anvil high-pressure (CA-HP) technique. To ensure phase purity and compositional uniformity, the high-purity precursors were employed, including samarium (Sm, 99.9%), iron (Fe, 99.99%), arsenic (As, 99.99%), iron (III) oxide ($Fe_2O_3$, 99%), and iron (II) fluoride ($FeF_2$, 99%). Due to the high volatility and reactivity of elemental arsenic, a preliminary synthesis of SmAs was conducted. Stoichiometric proportions of Sm and As were finely ground, compacted into pellets, which were then enclosed within a tantalum tube. This tube was then sealed in an evacuated quartz ampoule, which was heated at 550 °C for 15 hours, resulting in the formation of the SmAs phase. The resulting SmAs was then mixed with Fe, $Fe_2O_3$, and $FeF_2$ according to the stoichiometric chemical formula of $SmFeAsO_{1-x}F_x$. For each fluorine substitution content, approximately 0.45 grams of the mixed powder was prepared. These mixtures were pressed into disc-shaped pellets (diameter~6 mm) under a uniaxial pressure of 200 bar. The compacted pellets were placed into a boron nitride (BN) crucible. This crucible was placed within a pyrophyllite-based pressure medium to ensure efficient and uniform transmission of the high pressure to the sample during the synthesis process [37, 34]. A graphite tube served as the heating element in the high-pressure assembly. All high-pressure synthesis experiments were carried out under identical conditions at 1400 °C and 4 GPa (244 tonnes) for 1 h. First, the pressure was increased to 2 GPa (~150 tonnes) at a rate of 5 tonnes $min^{-1}$. The sample was then heated from room temperature to 1400 °C at a rate 15 °C/min, while simultaneously increasing the pressure at a rate of 5 tonnes $min^{-1}$ until the final pressure was reached. After reaching the target temperature and pressure, both were stabilized for approximately 5 min. Following the dwell time of 1 h, the samples were furnace-cooled naturally to room temperature while maintaining the applied pressure for an additional 5 min. Subsequently, the pressure was released at a rate of 7.5 tonnes $min^{-1}$ concurrently with furnace cooling. This well-controlled high-pressure sequence is essential for stabilizing the tetragonal Sm1111 phase and enabling fluorine incorporation well beyond the solubility limit achievable under ambient-pressure conditions. The resulting samples were characterized by X-ray diffraction and Raman spectroscopy for structural analysis, complemented by microstructural, transport, and magnetic measurements. Throughout this work, the fluorine substitution range is categorized into three regimes: underdoped ($0.05 \leq x < 0.2$), optimal doped ($0.2 \leq x \leq 0.25$), and overdoped ($0.25 < x \leq 0.40$).

X-ray diffraction (XRD) measurements were performed using a Rigaku SmartLab 3 kW diffractometer equipped with filtered Cu-$K_\alpha$ radiation ($\lambda$ = 1.5418 Å), operating at 30 mA and 40 kV. Data were recorded over a $2\theta$ range of 20°-70° with a step size of 0.02°/min using a Dtex250 linear detector. Lattice parameters were refined using BRUKER'S DIFFRAC.TOPAS software in conjunction with the ICDD PDF5+ 2024 database. Microstructural observations were conducted using an ultra-high-resolution nano-scanning electron microscope (nano-SEM), equipped with an energy-dispersive X-ray analysis (EDAX) system. Magnetic characterizations were performed using a vibrating sample magnetometer (VSM) integrated into a Physical Property Measurement System (PPMS). The magnetic susceptibility was measured under a constant magnetic field of 20 Oe across a temperature range of 5-60 K. Magnetisation hysteresis (*M-H*) measurements were also conducted at 5 K under the external magnetic fields of up to 9 T. The temperature-dependent electrical resistivity was measured using the standard four-probe technique within the PPMS under the applied magnetic fields up to 9 T. Raman spectroscopy was performed using a Horiba Jobin Yvon LabRam ARAMIS spectrometer, following the procedure detailed elsewhere [38]. The excitation source was a He-Ne ion laser emitting the visible light at the wavelength of 632.8 nm. The laser beam was focused to achieve a spot size of less than 3 mm on the specimen surface using a 100× objective lens (numerical aperture = 0.95), which also collected the backscattered radiation. A charge-coupled device (CCD) detector was used to detect the scattered light, which was dispersed by a diffraction grating of 2400 lines/mm. Stokes spectra were recorded within the 95-300 $cm^{-1}$ range with an acquisition time of 300 s for each spectrum. The laser power was reduced to approximately 140 μW to avoid local heating effects in the sample.

## Results:

Powder XRD patterns of $SmFeAsO_{1-x}F_x$ ($0.05 \leq x \leq 0.40$) bulks synthesized under high pressure are shown in Figure 1(a) along with the parent sample $x = 0$ prepared by CSP method.

All samples contain the tetragonal ZrCuSiAs-type Sm1111 phase (space group: *P4/nmm*) as the dominant phase, consistent with previous reports [39]. In addition to the main reflections, weak impurity peaks corresponding to SmOF, SmAs, and FeAs phases are observed, with their intensity increasing at higher nominal fluorine contents ($x \geq 0.25$). In the underdoped region ($x < 0.2$), small amounts of secondary phases, including SmOF and SmAs, are detected. For the optimal doped sample i.e., $x = 0.2$, the SmOF content is slightly increased compared to underdoped samples, while SmAs is nearly absent. However, for fluorine contents above 20% i.e., $x > 0.2$, the impurity phases SmOF and SmAs are increased rapidly, and a new FeAs phase is also detected, indicating that overdoping with fluorine contents promotes the formation of SmOF and segregation of SmAs/FeAs. A similar trend has been reported for CSP-synthesized $SmFeAsO_{1-x}F_x$, where lightly doped samples are nearly containing the target Sm1111 phase, but $x = 0.25$ contains significant impurities [5]. Notably, in the present study, we successfully achieved fluorine substitution up to 40% while retaining the tetragonal superconducting phase as the dominant phase. To the best of our knowledge, this is the first report of synthesizing Sm1111 with up to 40% fluorine substitution using a high-pressure growth method.

A magnified view of the main (102) reflection of the tetragonal superconducting phase is shown in Figure 1(b), revealing a systematic shift of the peak position toward higher $2\theta$ with increasing fluorine content. This behavior reflects a reduction of the interplanar spacing and a contraction of the crystal lattice due to fluorine substitution at the oxygen site. The lattice parameters $a$ and $c$, and the unit-cell volume $V$, were determined by least-squares Rietveld refinement using multiple reflections of the dominant Sm1111 phase and are presented in Figures 1(d)–1(f). The refinement profile for the $x = 0.05$ sample is shown in Figure 1(c), while those for $x = 0.25$ and 0.40 are provided in Supplementary Figure S1. For partially overlapping reflections, peak separation and deconvolution were applied prior to refinement to minimize the influence of impurity phases and ensure that the extracted lattice parameters correspond to the superconducting Sm1111 phase. For comparison, lattice parameters reported for F-doped $SmFeAsO_{1-x}F_x$ prepared by the CSP method [5] and for H-substituted $SmFeAsO_{1-x}H_x$ [25] are also included in Figures 1(d)–1(f). The *in-plane* lattice parameter $a$ decrease monotonically from 3.928(1) Å at $x = 0.05$ to 3.912(1) Å at $x = 0.30$, and remains nearly constant at 3.913(1) Å for $x = 0.40$ (Figure 1(d)). Similarly, the *out-of-plane* parameter $c$ shows a weak but nearly systematic contraction from 8.468(1) Å to 8.455(3) Å up to $x = 0.30$ and then saturates at higher nominal fluorine content (Figure 1(e)). The corresponding unit-cell volume decreases nearly linearly from 130.7(1) Å³ at $x = 0.05$ to 129.4(1) Å³ at $x = 0.30$ and remains essentially unchanged at 129.5(1) Å³ for $x = 0.40$ (Figure 1(f)). This lattice contraction is consistent with substitution of the slightly smaller $F^-$ ion (1.31 Å) for $O^{2-}$ (1.38 Å) in the SmO layer and follows approximately Vegard-like behavior [40] over most of the substitution range. Small deviations from perfect linearity, particularly in the overdoped regime, are observed and can be attributed to a combination of local lattice distortions induced by F-substitution, increasing fractions of secondary phases, and peak broadening and strain effects intrinsic to polycrystalline materials. Importantly, the evolution of the lattice parameters in the present F-substituted samples can be directly compared with that of H-substituted $SmFeAsO_{1-x}H_x$. In both systems, the lattice parameters $a$, $c$, and the unit-cell volume $V$ decrease approximately linearly with increasing substitution, indicating effective anion replacement and electron doping. However, as shown in Figures 1(d)–1(f), the lattice parameters and the unit-cell volume in H-doped SmFeAsO decrease more rapidly and linearly with hydrogen substitution ($x$) than in the F-doped system [5]. This reflects the much stronger chemical-pressure effect associated with $H^-$, whose effective size and bonding in the Sm(O,H) layer lead to a larger contraction of the lattice in addition to electron doping. In contrast, because the ionic radius of $F^-$ is closer to that of $O^{2-}$, the lattice contraction in $SmFeAsO_{1-x}F_x$ is more moderate and primarily reflects carrier doping rather than strong chemical pressure. As a result, F substitution provides a cleaner route to tune the electronic phase diagram of Sm-1111 while minimizing structural distortion compared to H substitution. A direct comparison with CSP-prepared F-doped samples further shows that the high-pressure-grown samples exhibit slightly smaller $a$, $c$, and $V$ at comparable nominal compositions, suggesting a somewhat higher effective fluorine incorporation under high-pressure synthesis conditions. At high nominal fluorine content ($x \geq 0.30$), impurity reflections become more prominent and the superconducting phase fraction decreases from ~82 % to ~60 %, a trend similar to that reported for CSP-prepared $SmFeAsO_{1-x}F_x$ at $x \approx 0.25$ [5]. Nevertheless, the persistence of lattice contraction and the consistency with Raman spectroscopy (Figure 3) confirm that fluorine substitution into the main tetragonal Sm1111 phase is achieved up to $x = 0.40$, even though impurity phases increasingly limit the phase purity in the overdoped regime.

Microstructural characterization of the synthesized $SmFeAsO_{1-x}F_x$ bulks was carried out on finely polished surfaces. Back-scattered electron (BSE) imaging was utilized to reveal both compositional contrast and microstructural morphology. The chemical nature of the different contrast regions observed in the BSE images was determined by energy-dispersive X-ray spectroscopy (EDX) point analysis and elemental mapping performed on representative areas. The corresponding EDX maps and spectra for the selected underdoped, optimal doped, and overdoped samples are provided in the Supplementary Information (Figure S2). These analyses confirm that the light-grey matrix corresponds to the SmFeAs(O,F) phase, while the bright and dark regions originate from Sm-rich ($Sm_2O_3$/SmOF) and As-rich (SmAs/FeAs) secondary phases, respectively. The selected BSE images corresponding to the underdoped, optimal doped, and overdoped compositions are displayed in Figures 2(a)-(b), 2(c)-(d), and 2(e)-(f), respectively. In these micrographs, bright areas represent the impurity $Sm_2O_3$ or SmOF phase, while the light grey matrix denotes the superconducting SmFeAs(O,F) phase, and the dark regions are normally attributed to pores. In some cases, the dark region can represent SmAs or FeAs as secondary phases. The underdoped sample ($x$ = 0.05) appears dense and largely containing the target phase, with only a few tiny bright inclusions. The optimal doped sample ($x$ = 0.20; Figure 2(c)) exhibits a more homogeneous and denser microstructure compared to the underdoped sample. A small number of impurity phases, identified as SmAs and $Sm_2O_3$, are observed at a few locations, suggesting that both homogeneity and sample density are improved at 20% fluorine substitution level. In the overdoped sample ($x$ = 0.40; Figure 2(e)), the secondary phases such as $Sm_2O_3$/SmOF become more abundant and coarser, many residing along grain boundaries and forming weak Josephson-junction-like morphologies, consistent with typical behavior in overdoped F-doped Sm1111 [41] and with XRD evidence of increased impurity formation at high fluorine content. High-magnification BSE images provide insight into grain-boundary (GB) quality. For $x$ = 0.05 (Figure 2(b)) and $x$ = 0.20 (Figure 2(d)), many grains are well-connected, and the grain boundaries are clean and continuous, with no observable impurity phases at the nanoscale. This indicates a good electrical percolation path of the current in these bulks. For $x$ = 0.40 (Figure 2(f)), the boundary network remains largely continuous; however, faint nanometer-scale contrasts appear more frequently at some interfaces, consistent with the presence of thin intergranular secondary phases. Overall, Figure 2 demonstrates that the high-pressure synthesis produces ~98% dense (relative sample density) compacts with well-connected grain boundaries across the substitution series. The gradual increase of $Sm_2O_3$/SmAs/SmOF content from optimal to overdoped compositions, as also observed in XRD, confirms the progressive formation of impurity phases at high fluorine substitution.

Raman spectra were collected at room temperature directly from the as-prepared polycrystalline samples, without any post-synthesis polishing or surface modification. Due to the polycrystalline morphology, noticeable variations were observed in the relative intensities of the spectral features, depending on the measurement location. This variability is likely associated with differences in grain orientation and local crystallographic alignment at the measurement sites. As a result, we did not pursue a detailed analysis of intensity trends related to fluorine substitution, since such variations could not be unambiguously attributed to compositional effects alone. To account for statistical variation and ensure consistency in peak position identification, Raman measurements were performed at three distinct locations on each sample for a given fluorine concentration. The acquired spectra were then processed using peak deconvolution with Lorentzian functions, allowing for the accurate extraction of vibrational mode positions. For all prepared $SmFeAsO_{1-x}F_x$ samples, the representative spectra from the underdoped, optimal doped, and overdoped regions are shown in Figure 3(a). The spectra reveal signals corresponding to lattice vibrations (phonons) of Sm, As, and Fe atoms, with $A_{1g}$, $A_{1g}$, and $B_{1g}$ symmetries, respectively, consistent with previous reports on pristine SmFeAsO, Cu-doped material ($SmFe_{1-x}Cu_xAsO$) and other similar *RE*-based oxypnictides [42, 43, 44, 38, 45]. Across the entire substitution range ($x = 0.05 – 0.40$), including the overdoped region ($x = 0.30$–0.40), the overall Raman spectra confirm that the tetragonal Sm1111 structure is preserved, indicating that the chemical bonding remains stable upon fluorine substitution, as depicted in Figure 3(b). Analysis of the phonon frequencies reveals systematic shifts with increasing fluorine content: the Sm-related mode softens by ~8 $cm^{-1}$, while the Fe-related mode hardens by ~4 $cm^{-1}$, whereas the As-related mode shows only minor changes within experimental error (Figure 3b). These continuous and monotonic shifts across the full substitution range provide phase-sensitive evidence of successful fluorine incorporation, even in nominally overdoped samples. The observed trends reflect subtle modifications in the local bonding environment and electronic structure induced by electron doping through F substitution, consistent with previous observations in $NdFeAsO_{1-x}F_x$ ($x$ = 0-0.2) [45]. Together with the XRD data, these results confirm that high-pressure synthesis enables effective fluorine substitution up to $x = 0.40$, despite the presence of minor impurity phases in the overdoped regime, and that the main Sm1111 phase remains the dominant superconducting phase.

The temperature-dependent resistivity of all prepared $SmFeAsO_{1-x}F_x$ bulks ($0.05 \leq x \leq 0.40$) measured under zero magnetic field is shown in Figure 4(a) up to room temperature. All compositions exhibit metallic behaviour from 300 K down to the superconducting transition, with no pronounced low-temperature upturn, indicating that charge transport is not dominated by carrier localization or magnetic scattering as reported for the parent SmFeAsO [5]. The samples $x$ = 0.05, 0.10, 0.20 display only slight variations in the normal-state resistivity across the measured range. However, in the overdoped region ($x \geq 0.25$), XRD and SEM analyses reveal an increasing amount of secondary phases, some of which (e.g., FeAs, SmAs) are metallic. These impurity phases can therefore contribute to the measured resistivity, particularly to its absolute magnitude in the normal state. Consequently, the observed reduction of resistivity with increasing $x$ above 0.20 cannot be attributed solely to electron doping of the Sm1111 phase, but is likely influenced by a combination of increased carrier concentration, improved densification in the CA-HP synthesized bulks, and parallel conduction through metallic impurity phases.

The low-temperature resistivity of the samples ($0.05 \leq x \leq 0.40$) is presented in Figure 4(b). The superconducting transition temperatures are defined using 90% and 10% of the normal-state resistivity for the onset transition temperature $T_c^{onset}$ and offset transition temperature $T_c^{offset}$, respectively. The underdoped samples show onset transition temperatures of ~54.4 K and 55.6 K for $x$ = 0.05 and 0.10, respectively, which are significantly ~10-17 K higher than those typically reported for F-doped Sm1111 prepared by the CSP method [5]. For compositions in the optimal ($0.20 \leq x \leq 0.25$) and overdoped ($0.25 < x \leq 0.40$) regions, the $T_c^{onset}$ remains nearly constant at about ~55-57 K, in agreement with previous reports for $SmFeAsO_{1-x}F_x$ [5]. While the normal-state resistivity in the overdoped region may be affected by metallic secondary phases, the persistence of sharp superconducting transitions and a clear zero-resistance state up to $x$ = 0.40 demonstrates that superconductivity originates from the $SmFeAsO_{1-x}F_x$ phase. The slightly reduced $T_c^{onset}$ and moderate broadening of the transition for $x$ = 0.40 are consistent with the increased fraction of impurity phases inferred from XRD and SEM, which can introduce percolative current paths and reduce the effective superconducting volume fraction. Accordingly, a slightly larger resistive broadening is observed under different current measurements ($I$ = 5, 10 and 20 mA) for the overdoped samples, as shown in the Supplementary Figure S3. Nevertheless, the systematic evolution of the superconducting transition with fluorine content supports the conclusion that fluorine substitution into the Sm1111 lattice is maintained even in the overdoped regime.

To further evaluate sample quality, we extracted the onset $T_c^{onset}$ and offset $T_c^{offset}$ superconducting transition temperature, transition width $\Delta T = T_c^{onset} - T_c^{offset}$, room-temperature resistivity $\rho_{300K}$, and residual resistivity ratio $RRR$ (= $\rho_{300K} / \rho_{60K}$), as shown in Figures 4(c)-(f) respectively. For polycrystalline $SmFeAsO_{1-x}F_x$ samples, the $RRR$ is affected by grain boundaries, porosity, and secondary phases; therefore, it cannot be interpreted as an intrinsic measure of the sample quality. Here, $RRR$ is used only as a relative indicator of the evolution of charge transport and scattering with fluorine substitution for samples prepared under identical high-pressure growth conditions. Figure 4(c) shows the variation of $T_c^{onset}$ and $T_c^{offset}$ as a function of nominal fluorine contents ($x$). The onset temperature increases from 54.4 K at $x$ = 0.05 to a maximum of 57.3 K at $x$ = 0.20, indicating the approach to optimal fluorine substitution. Upon further increasing $x$ into the nominally overdoped region, $T_c^{onset}$ decreases slightly to 56.8 K ($x$ = 0.25), 56.3 K ($x$ = 0.30), and 55.5 K ($x$ = 0.40). A similar evolution is observed for $T_c^{offset}$. While such a weak suppression of superconducting transition beyond the optimal doping is consistent with the expected behavior of electron-doped Sm1111, structural and microstructural analyses show that impurity phases (SmOF, SmAs, and FeAs) become increasingly abundant for $x \geq 0.25$. These metallic and insulating secondary phases can reduce the effective superconducting volume fraction and introduce weak links between grains. The transition width $\Delta T$ is large for underdoped samples, measuring 7.7 K at $x$ = 0.05 and 8.5 K at $x$ = 0.10. However, it decreases nearly two- to threefold for the optimal doped sample, which has a transition width of 3 K at $x$ = 0.20, reflecting improvements in sample density, intergrain connectivity, and a decrease in impurity content. For higher substitution, $\Delta T$ is 2.7 K at $x$ = 0.25 and slightly increases in the overdoped region (3.4 K at $x$ = 0.30 and 0.40), consistent with the formation of secondary phases. Therefore, the observed reduction of $T_c^{onset}$ and the broadening of the transition in the overdoped region likely reflect a combination of intrinsic overdoping effects and extrinsic degradation due to impurity formation, rather than a purely electronic suppression of superconductivity. The room-temperature resistivity $\rho_{300K}$ (Figure 4(e)) is ~1.8 mΩ·cm for underdoped and optimal doped ($x$ = 0.2) samples, and decreases monotonically for higher fluorine contents ($x$ = 0.25–0.40), suggesting increased carrier density and fewer effective scattering centres in these dense polycrystalline samples. The *RRR* (Figure 4(f)) shows a dome-like dependence on fluorine content: it rises rapidly with substitution in the underdoped region, reaching a maximum of ~5.75 at $x$ = 0.20, then decreases to 4.97 ($x$ = 0.25), 4.65 ($x$ = 0.30), and 3.82 ($x$ = 0.40), indicating weaker homogeneity and intergrain connectivity in both underdoped and overdoped samples, likely due to the presence of the secondary phases as observed in the structural and microstructural analysis. Overall, this analysis indicates that a high $T_c^{onset}$ along with a small $\Delta T$ and a large *RRR*, is achieved within the optimal fluorine substitution range $x \approx 0.20$–$0.25$, where the dense and homogeneous F-doped Sm1111 can be obtained using the present high-pressure technique. The combination of resistivity and structural analysis suggests that $T_c^{onset}$ generally increases as XRD peaks shift to higher $2\theta$ values and the lattice constants contract with increasing fluorine content (Figures 1(b)-(e)). However, the continuous increase of $T_c^{onset}$ is not observed (Figure 4(c)) with fluorine substitution, even with the steady contraction of the lattice parameters (Figure 1(c) and (d)). Fluorine substitution reduces the lattice size and injects electrons, and the observed $T_c^{onset}$ exhibits a dome-shaped relationship with carrier concentration. We expect that near optimal substitution ($x \approx 0.20$-$0.25$), the FeAs bands are tuned to maximize $T_c^{onset}$, while further fluorine addition moves the system into the overdoped regime, where $T_c^{onset}$ plateaus or declines. Simultaneously, higher fluorine substitution promotes formation of secondary SmOF/SmAs/FeAs phases, as indicated by diffraction and scanning electron microscopy (SEM), and probably contribute to pair-breaking and intergranular scattering. These secondary phases reduce the effective superconducting cross-section and introduce weak links, consistent with the modest $\Delta T$ broadening and $RRR$ reduction for $x \geq 0.3$. Therefore, the observed plateau in $T_c^{onset}$, despite ongoing lattice contraction, likely results from surpassing the electronic optimum, while extrinsic scattering progressively increases, explaining a possibility why $T_c^{onset}$ does not rise further with fluorine content in these F-doped SmFeAsO samples prepared by the CA-HP technique.

The variation of electrical resistivity ($\rho$) with temperature under the applied magnetic field up to 9 T was presented in Figure 5(a) for three fluorine compositions: $x = 0.05$, 0.25, and 0.40, corresponding to the underdoped, optimal doped and overdoped regimes, respectively. These measurements were conducted under magnetic fields of up to 9 T within a temperature range of below 70 K. In the absence of an external field, each sample exhibit sharp superconducting transitions together with metallic behaviour in the normal state above the $T_c$, reflecting good sample homogeneity and strong intergrain connectivity, consistent with the SEM observations. With increasing magnetic field, the superconducting transitions shift systematically toward lower temperatures. However, the transition broadening is significantly weaker for the optimal doped sample ($x \approx 0.25$) than for the underdoped and overdoped compositions. This reflects the evolution of grain connectivity and phase purity with fluorine substitution: near optimal doping, grains are better coupled and impurity phases are minimized, whereas at the edges of the superconducting dome ($x = 0.05$ and especially $x = 0.40$), non-superconducting inclusions and weak links lead to a distribution of local $T_c$ values and vortex pinning energies, producing broader resistive transitions under magnetic field. To quantify the superconducting phase boundaries, we employed standard resistive criteria relative to the normal-state resistivity $\rho_n(T)$: the field at which $\rho(T,H) = 0.9\,\rho_n(T)$ defines the upper critical field $H_{c2}(T)$, while the field at which $\rho(T, H) = 0.1\rho_n(T)$ defines the irreversibility field $H_{irr}(T)$ (corresponding to the loss of macroscopic dissipation-free transport). Using these criteria, the $H_{c2}(T)$ and $H_{irr}(T)$ phase diagrams are constructed and are shown in the inset of Figures 5(a)-(c) for the respective samples. From the data, the slopes $(dH_{c2}/dT)|_{Tc}$ is determined as -17.3, -5.7, and -3.2 T $K^{-1}$ for the $x$ = 0.05, 0.25, and 0.40 samples, respectively. The upper critical field $H_{c2}(0)$ is then estimated using the Werthamer–Helfand–Hohenberg (WHH) single-band orbital formula [46], $\mu_0 H_{c2}(0) \approx -0.693\ T_c\ (dH_{c2}\ /\ dT)|_{T=Tc}$. Taking the resistive onset transition temperatures $T_c^{onset}$ from transport measurements at zero magnetic field (54.4 K, 56.8 K, and 55.5 K for $x$ = 0.05, 0.25, and 0.40, respectively; Figure 4(c)), we obtain zero-temperature upper critical fields of $\mu_0 H_{c2}(0) \approx 223$ T for $x$ = 0.25 and $\mu_0 H_{c2}(0) \approx 125$ T for $x$ = 0.40. The underdoped sample exhibits an even steeper initial slope, yielding a significantly larger orbital-limited $H_{c2}(0)$, consistent with earlier reports on F-doped Sm-1111 single crystals, where unusually large initial slopes and strong paramagnetic limitation were also observed [6]. These WHH-derived values are considerably higher than directly measured $H_{c2}(0)$ values reported for Sm1111 single crystals and thin films. For example, Lee *et al.* [47] reported $\mu_0 H_{c2}$ values of ~65–70 T in $SmFeAsO_{1-x}F_x$ single crystals, while recent high-quality Sm1111 thin films exhibit $\mu_0 H_{c2}$ values in the range of 80-100 T depending on field orientation [48]. The larger WHH-extrapolated values obtained in our CA-HP polycrystalline samples reflect the well-known tendency of the WHH model to overestimate $H_{c2}$ in FBS, because it neglects Pauli pair-breaking and multiband effects. The corresponding Ginzburg–Landau coherence lengths were estimated from $\mu_0 H_{c2}(0)$ using $\xi(0) = \sqrt{[\Phi_0/(2\pi\ \mu_0 Hc_2(0))]}$ [49], where $\Phi_0$ is the flux quantum. The calculated values are $\xi(0) \approx 0.71$ nm ($x$ = 0.05), 1.21 nm ($x$ = 0.25), and 1.71 nm ($x$ = 0.40). These values are comparable to those reported for Sm1111 single crystals and thin films, which typically fall in the range 1–2 nm [48, 47], confirming that superconductivity in the present samples remains in the strongly type-II regime despite the presence of impurity phases. For clarity, a comparative summary of $H_{c2}$ calculated by WHH model and the corresponding coherence lengths for polycrystalline F-doped SmFeAsO, single crystals, and thin films is provided in Table 1.

It is well established that high-$T_c$ iron-based superconductors are multiband in nature, and their upper critical field behavior can be understood within the framework of multiband s± pairing [50]. While the WHH model considers only orbital pair-breaking, real superconductors are strongly influenced by both orbital and spin paramagnetic effects. To evaluate the relative importance of orbital and spin-paramagnetic pair breaking, we estimated the Maki parameter [51], $\alpha = 1.44 H_{c2}^{orbital}(0)\ /\ H_{c2}^{para}(0)$, where the spin paramagnetic limiting field is given by $H_{c2}^{para}(0) = 1.84\ T_c$ according to BCS theory [52]. For our samples, the large WHH-derived orbital fields lead to α > 1, indicating that Pauli paramagnetic effects are significant. It suggests the possibility of a Fulde–Ferrell–Larkin–Ovchinnikov (FFLO) state [53] and is also reported that in multiband superconductors, both $H_{c2}^{orbital}$ and the stability of the FFLO state can be tuned by the doping type and level. This may explain the distinct $H_{c2}(0)$ values as we observe across underdoped, optimal doped, and overdoped F-doped Sm1111 samples. Similar large Maki parameters (α > 1) have been reported for F substituted Sm1111 single crystals and H-substituted 1111 thin films [47, 48] as summarized in Table 1. In all these cases, the upper critical field behavior is strongly influenced by paramagnetic limiting and multiband superconductivity effects. In H-substituted LaFeAsO systems [22], the expanded substitution range enables access to a second superconducting dome accompanied by enhanced effective masses and stronger Pauli limitation, leading to large Maki parameters and complex $H_{c2}(T)$ behavior. Compared with H substitution, fluorine substitution introduces a smaller chemical-pressure effect, making the present CA-HP processed F-doped Sm1111 samples particularly valuable for isolating the role of carrier doping on $H_{c2}$ and paramagnetic effects. The high Maki parameters observed here suggest that even in polycrystalline, impurity-containing samples, the intrinsic superconducting state retains the strong Pauli-limited and multiband character of Sm1111. It should be noted that the extremely high $\mu_0 H_{c2}$ values obtained from WHH extrapolation represent orbital-limited upper bounds rather than physically attainable critical fields, as direct high-field measurements in 1111 systems typically saturate near 80–100 T [47, 48]. Nevertheless, the large initial slopes and short coherence lengths demonstrate that the CA-HP synthesized $SmFeAsO_{1-x}F_x$ samples possess robust intrinsic superconductivity comparable to the best Sm1111 single crystals and thin films.

The magnetic-field-induced broadening of the superconducting transition originates from vortex motion in the mixed state. Although the samples were synthesized under high pressure, X-ray diffraction and microstructural analysis show no evidence of preferred crystallographic orientation; the pellets can therefore be regarded as randomly oriented polycrystalline bulks. Consequently, the magnetic field (applied perpendicular to the current and to the pellet surface) probes an orientation-averaged vortex response, and the extracted thermally activated flux flow (TAFF) parameters represent the effective, macroscopic pinning landscape rather than a specific crystallographic direction. In polycrystalline FBS, dissipation is commonly governed by TAFF [54], where vortices overcome pinning barriers via thermal activation. In this regime, the resistivity can be expressed as: $\rho(T, H) = \rho_0 \exp[-U_0(H) / k_B T]$ with $U_0$ as the effective activation energy. To minimize sample-to-sample variations, the resistivity was normalized by the normal-state resistivity ($\rho_n$) just above the superconducting transition. Accordingly, Arrhenius plots of ln($\rho$ / $\rho_n$)) versus 1/$T$ display straight-line behavior in the TAFF regime, where the slope $-U_0(H)/k_B$ provide the activation energy $U_0(H)$ with different magnetic fields. Representative Arrhenius plots for the samples with $x$ = 0.05, 0.25, and 0.40 under fields up to 9 T are shown in Figures 6(a)-(c). Quasi-linear behavior is observed in a narrow temperature interval below $T_c$, indicating that vortex motion in this region is governed by thermally activated hopping. With increasing magnetic field, the slope decreases systematically, reflecting the reduction of the activation energy $U_0$ as the vortex density increases. The field dependence of the activation energy $U_0(H)$, extracted from these slopes is summarized in Figure 6(d). For all compositions, the field dependence of the activation energy $U_0(H)$ follows a power law, $U_0 \propto H^{-\eta}$, which is characteristic of collective vortex pinning in FBS. The underdoped ($x$ = 0.05) and optimally doped ($x$ = 0.25) samples exhibit a relatively weak field dependence ($\eta \approx 0.29$, $\eta \approx 0.38$) at low fields and a stronger suppression ($\eta \approx 0.50$, $\eta \approx 0.44$) of $U_0$ at higher fields, consistent with a crossover from single-vortex or small-bundle pinning to more collective vortex motion [55]. In contrast, the overdoped sample ($x$ = 0.40) shows significantly smaller activation energies and a more rapid decrease of $U_0$ with magnetic field. This behavior indicates a weaker effective pinning landscape and enhanced vortex mobility. Importantly, this reduction of $U_0$ cannot be attributed solely to intrinsic overdoping effects. XRD and microstructural analysis reveal an increased fraction of metallic secondary phases (such as SmAs and FeAs) at high nominal fluorine content, which can introduce parallel conductive paths and weaken the effective superconducting cross section. These impurity phases are therefore expected to contribute to the reduced activation energy and the enhanced dissipation observed in the overdoped samples. Furthermore, this evolution of $\eta$ is consistent with reports on other FBS (e.g., Nd1111, Sm1111, FeSe) [55] [56]. In our polycrystalline bulks, the absolute values of $U_0$ are smaller than those reported for F-doped Sm1111 single crystals [6] because grain boundaries and secondary phases introduce weak links and limit the macroscopic pinning force. Thus, the evolution of $U_0(H)$ across the fluorine- substitution series reflects both intrinsic changes in vortex pinning and extrinsic effects arising from microstructural disorder and impurity phases, which become increasingly important in the overdoped regime. In addition, we also attempted to extract $U_0$ using a broader fitting range following the commonly adopted practical procedure proposed by Zhang *et al.* [57]. However, for the present polycrystalline samples, especially in the overdoped region, the resistive transitions are affected by grain-boundary and secondary-phase contributions, leading to unstable and non-physical fitting results. Therefore, only the narrow-temperature Arrhenius analysis close to $T_c$, where linear behavior is clearly observed, is used in this work.

The magnetic susceptibility of all prepared samples is measured in zero-field-cooled (ZFC) and field-cooled (FC) modes under an applied field of 20 Oe in the temperature range of 5–60 K, and the normalized magnetization as a function of temperature is presented in Figure 7(a). All $SmFeAsO_{1-x}F_x$ bulks exhibit a clear diamagnetic response for each composition, confirming bulk superconductivity in these polycrystalline samples. Normalization of the magnetization enables direct comparison of the superconducting transition behavior across different fluorine substitution levels. The underdoped samples ($x$ = 0.05 and 0.10) display the broadest transitions in the ZFC mode along with a pronounced ZFC-FC separation extending well below $T_c$. The transition temperatures are observed at approximately 50 K for $x$ = 0.05 and 51 K for $x$ = 0.10. The broad transitions and the presence of a double-step feature suggest weaker intergranular coupling and a wider distribution of local superconducting transition temperatures. In contrast, the $x$ = 0.20 sample, which belongs to the optimally doped region, exhibits an onset transition at 56 K with a clear double-step feature. The $x$ = 0.25 sample exhibits an almost single-step transition with an onset at 55 K, indicative of improved grain connectivity compared with the $x$ = 0.05, 0.10, and 0.20 samples. With further fluorine substitution ($x$ = 0.30 and 0.40), the transitions broaden again and display double-step features, although the onset temperature remains around 54 K. Interestingly, the higher-doped $x$ = 0.40 sample still develops a robust diamagnetic state, albeit with a slightly lower onset temperature and modest broadening compared to the sample $x$≈0.25. The observed transition temperatures are ~1 K lower than those determined from transport measurements, as discussed above. Slightly negative FC branches are observed for all fluorine-doped samples, reflecting strong vortex pinning and irreversible flux trapping [38]. The maximum transition temperature for $x$ ≈ 0.20–0.25 coincides with the narrowest resistive transitions and the highest residual resistivity ratio (*RRR*). In contrast, the highest fluorine-substituted sample ($x$ = 0.40) shows a slightly reduced $T_c$ and broader transitions, likely due to the presence of secondary $SmAs/Sm_2O_3$ phases, as revealed by structural and microstructural analysis. These non-superconducting inclusions reduce the effective cross-sectional area and introduce weak links at some grain boundaries, even though the pellets remain overall dense and well-connected.

A high critical current density ($J_c$) is essential for the technological application of a superconducting material. To evaluate this parameter, magnetic hysteresis (*M-H*) measurements were conducted at 5 K under the magnetic fields up to 9 T for all prepared bulks. The hysteresis width ($\Delta m$), defined as the difference between the magnetic moments recorded during the increasing and decreasing field cycles, was used to estimate $J_c$ according to the extended Bean model: $J_c = 20\Delta m/Va(1-a/3b)$, where $a$ and $b$ denote the in-plane dimensions of the sample ($a < b$), and $V$ represents the sample volume [58]. The calculated $J_c$ at 5 K as a function of applied magnetic field is shown in Figure 7(b) for samples with $x = 0.05$, 0.10, 0.20, 0.25, 0.3 and 0.4. The sample with $x = 0.05$ exhibits high $J_c$ values at low fields, which decrease rapidly with increasing magnetic field and reach approximately $10^3$ A/cm$^2$ at 3 T, remaining nearly constant up to 9 T. Overall, the $J_c$ of the $x = 0.05$ sample remains around $3 \times 10^3$ A/cm$^2$ across the entire magnetic field range, which is higher than that reported for F-doped Sm-1111 prepared by the CSP method [5]. The behavior of the $x = 0.10$ sample is similar to that of $x = 0.05$, with $J_c$ values increased slightly throughout the entire magnetic field range. With further fluorine substitution ($x = 0.20$), the $J_c$ value and its field dependence remain nearly the same as for the $x = 0.10$ sample. A significant enhancement is observed for the $x = 0.25$ sample, where $J_c$ increases by approximately an order of magnitude at low magnetic fields, though it decreases by about 4000 A/cm$^2$ under fields up to 9 T. In contrast, the $x = 0.30$ and $x = 0.40$ samples, which lie in the overdoped region, exhibit reduced $J_c$ compared with $x = 0.25$. Notably, these overdoped samples also exhibit a significantly faster degradation of $J_c$ with the magnetic field, likely due to the presence of impurity phases, as corroborated by the XRD and transport measurements discussed above. Interestingly, the $J_c$ values of the samples prepared by the high-pressure method exhibit robust behavior against magnetic fields, which is highly favorable for magnetic applications. The $x = 0.40$ sample shows the fastest suppression of $J_c$ with increasing field, attributable to the larger population of SmAs/$Sm_2O_3$ impurity phases decorating the grain boundaries and forming Josephson junctions. These features, directly observed in SEM images, disrupt the superconducting network and reduce the effective intergranular critical current under high Lorentz forces. To gain further insight into the $J_c$ performance, the flux pinning force ($F_p$) was calculated using the relation $F_p = \mu_0 H \times J_c$ [59], as shown in the inset of Figure 7(b). The analysis confirms an enhanced pinning force for the $x = 0.25$ sample compared with the other compositions, indicating that increased flux pinning contributes to its superior $J_c$. Overall, the obtained $J_c$ values are an order of magnitude higher, with improved field robustness, compared with those reported for F-doped Sm1111 synthesized by the CSP method.

## Discussion:

The composition dependence of the superconducting properties of $SmFeAsO_{1-x}F_x$ prepared via the CA-HP route is summarized in Figures 8(a) and 8(b), showing the superconducting onset transition $T_c$ from transport measurements and the critical current density $J_c$ from magnetic measurements, respectively. For comparison, $T_c$ and $J_c$ data for F-doped Sm1111 synthesized by the CSP method, reported by Singh *et al.* [5] and Wang *et al.* [26] are also included in Figure 8 to evaluate the effects of different growth conditions. In addition, the onset superconducting transition data for H-doped SmFeAsO are plotted in Figure 8(a) for further comparison. Notably, Wang *et al* employed a high-temperature (~1200°C) synthesis process of F-doped Sm1111, whereas Singh *et al.* used a low-temperature (~900°C) CSP route. A clear synthesis-temperature effect is observed: the high-temperature CSP samples exhibit significantly lower $T_c$ than the low-temperature CSP samples at the same nominal fluorine content, consistent with partial fluorine loss during high-temperature synthesis. This demonstrates that suppressing fluorine evaporation is crucial for achieving effective electron substitution and high $T_c$. Compared with both CSP routes, CA-HP–synthesized $SmFeAsO_{1-x}F_x$ exhibits a pronounced enhancement of $T_c$ in the underdoped regime. In particular, $T_c$ is higher by approximately 17 K at $x = 0.05$ and 13 K at $x = 0.10$ compared with CSP samples. In the optimal doped region ($x \approx 0.20–0.25$), $T_c$ reaches ~56–57 K and becomes comparable for CA-HP and low-temperature CSP samples, indicating that all methods can approach optimal carrier concentration when fluorine incorporation is sufficient. In the overdoped region ($x > 0.25$), $T_c$ of CA-HP samples decreases slightly by about 1–2 K relative to CSP samples, but remains nearly constant at ~55–57 K over the wide range $0.20 \leq x \leq 0.40$. Structural analysis provides insight into these trends: CA-HP samples exhibit slightly smaller lattice parameters and unit-cell volumes than CSP-processed samples at the same nominal fluorine content (Figure 1), indicating more effective fluorine incorporation and enhanced densification under high-pressure synthesis. However, in the overdoped region, secondary phases such as SmOF, SmAs, and FeAs increasingly segregate at grain boundaries, reducing the superconducting volume fraction and weakening intergranular coupling. The combined effects of overdoping, lattice detuning, and impurity-rich grain boundaries likely account for the modest reduction of $T_c$ at high fluorine content [4]. Interestingly, the evolution of $T_c$ in H-doped $SmFeAsO_{1-x}H_x$ closely follows that of low-temperature CSP-processed F-doped samples in the underdoped and optimal doped regions. However, in the overdoped regime, $T_c$ in H-doped Sm1111 decreases much more rapidly, whereas F-doped Sm1111 synthesized by either the low-temperature CSP or CA-HP routes shows only a weak suppression of $T_c$. This difference highlights the stronger structural and electronic perturbation introduced by H substitution compared with fluorine, and underscores the advantage of fluorine substitution—especially when combined with CA-HP synthesis—for stabilizing high-$T_c$ superconductivity over a broad doping range.

Figure 8(b) presents the critical current density ($J_c$) at 5 K as a function of nominal fluorine composition ($x$) for CA-HP samples, alongside previous CSP-based reports [5] [26]. F-doped samples prepared by low-temperature CSP exhibited maximum $J_c$ values on the order of $10^3$ A/cm² in the underdoped and optimal doped regions [5], whereas high-temperature CSP samples reached ~$10^4$ A/cm² at optimal doping ($x$ = 0.20) [5], with a slight decrease in the overdoped region. For high-pressure synthesized samples, $J_c$ exhibits a dome-like behavior, reaching a maximum at $x$ = 0.25 in the optimal doped region. In the underdoped region, $J_c$ is slightly lower but remains higher than that of CSP-processed samples (Figure 8(b)), limited primarily by reduced carrier density and weaker intergranular coupling. In the overdoped region, $J_c$ decreases due to the increasing presence of impurity phases such as SmAs and $Sm_2O_3$, as well as the formation of weak-link Josephson junctions at grain boundaries, which disrupt the current percolation pathways and accelerate the field-induced decay of $J_c(H)$ in the samples grown under HP conditions. Compared with CSP-processed samples, our HP-grown samples achieve $J_c$ values comparable to high-temperature CSP reports at equivalent fluorine substitution levels. The phase diagram in Figure 8 highlights a practical processing window for F-doped SmOFeAs at $x \approx$ 0.25-0.30, where both $T_c^{onset}$ and $J_c$ are simultaneously optimized through HP growth process. Overall, high-pressure synthesis significantly enhances the superconducting properties across the entire underdoped region by stabilizing fluorine in the lattice and improving densification, whereas the overdoped side remains limited by inhomogeneity and the presence of impurity phases. These trends suggest that $T_c$ is governed by the interplay between carrier concentration and lattice geometry, while the $J_c$ dome is controlled by flux pinning, which depends on phase purity and grain connectivity. Together, these insights provide a coherent framework for optimizing Sm1111 bulk materials for high-current applications.

Overall, these studies indicate that HP synthesis is a highly effective approach for underdoped $SmFeAsO_{1-x}F_x$ samples, enhancing the superconducting onset transition temperature by up to ~17 K and increasing $J_c$ by nearly an order of magnitude compared with CSP method [5]. The HP growth route simultaneously improves both the superconducting transition and the critical current properties, whereas the CSP process typically optimizes only one parameter per synthesis route. For instance, low-temperature CSP synthesis enhances the superconducting transition temperature but yields a lower $J_c$ (~$10^3$ A/cm²), while high-temperature CSP synthesis increases $J_c$ by an order of magnitude (~$10^4$ A/cm²) at the expense of reducing $T_c$ by ~10 K. Our HP synthesis approach therefore offers a unique advantage by enabling simultaneous control of superconducting transition and current-carrying capability, while also extending the fluorine-substitution-dependent phase diagram, highlighting its potential for both fundamental and applied research. Nevertheless, further optimization and detailed investigations are needed to fully exploit this approach for practical high-current devices.

# Conclusion:

Polycrystalline $SmFeAsO_{1-x}F_x$ bulks with $0.05 \leq x \leq 0.40$ were successfully synthesized using an *in-situ* CA-HP method and comprehensively characterized. The tetragonal Sm1111 phase (*P4/nmm*) is preserved across the entire substitution range, accompanied by a systematic contraction of the lattice parameters and unit-cell volume that confirms effective fluorine incorporation, further supported by Raman spectroscopy measurements. Compared with conventionally prepared CSP samples, CA-HP synthesis results in slightly smaller lattice dimensions, indicating improved fluorine incorporation and enhanced densification. High-resolution backscattered electron imaging shows well-connected grains in the optimal-doped samples, while the overdoped region exhibits secondary phases (SmOF, $Sm_2O_3$, and SmAs) along grain boundaries, which may weaken intergranular coupling. Transport and magnetization measurements confirmed bulk superconductivity across the substitution range up to $x = 0.4$, with a dome-like dependence of $T_c$ and $J_c$ on fluorine content. The optimal doped compositions ($x \approx 0.20$-$0.25$) exhibited the highest performance, achieving $T_c$ of ~55-57 K and $J_c$~$10^4$ A cm$^{-2}$, with nearly field-independence $J_c$ at high fields. Analysis of the upper critical field $H_{c2}(0)$ indicated the dominance of the Pauli paramagnetic effect, consistent with short coherence lengths (~1–2 nm) and the multiband superconductivity in Sm1111. Thermally activated flux-flow behavior revealed a power-law field dependence of the activation energy, $U_0(H) \propto H^{-\eta}$, across all substitution regions, suggesting collective vortex pinning at high magnetic fields (>3 T). These results demonstrate that high-pressure synthesis effectively stabilizes dense Sm1111 bulks with enhanced superconducting properties and also extends the accessible superconducting phase diagram, suggesting their potential for practical wire and tape applications.

## CRediT authorship contribution statement

**Mohammad Azam:** Writing – review & editing, Writing – original draft, Investigation, Formal analysis, Data curation. **Tatiana Zajarniuk:** Data curation, Investigation, Resources, Writing – review & editing. **Ryszard Diduszko:** Formal analysis, Investigation, Resources, Data curation, Writing – review & editing. **Taras Palasyuk:** Writing – review & editing, Formal analysis, Investigation, Data curation. **Cezariusz Jastrzębski:** Writing – review & editing, Resources, Data curation. **Andrzej Szewczyk:** Writing–review & editing, Resources, Data curation. **Hiraku Ogino:** Writing–review & editing, Investigation, Resources, Methodology. **Shiv J. Singh:** Writing–review & editing, Writing – original draft, Visualization, Validation, Supervision, Software, Resources, Methodology, Investigation, Funding acquisition, Formal analysis, Conceptualization.

## Declaration of competing interest

The authors declare that they have no known competing financial interests or personal relationships that could have appeared to influence the work reported in this paper.

## Data availability

The raw/processed data required to reproduce these findings cannot be shared at this time due to technical or time limitations. Data are available upon request to the corresponding author.

**Acknowledgments:**

The work was funded by SONATA-BIS 11 project (Registration number: 2021/42/E/ST5/00262) sponsored by National Science Centre (NCN), Poland. SJS acknowledges financial support from National Science Centre (NCN), Poland through research Project number: 2021/42/E/ST5/00262. We would like to acknowledge H. Kito, AIST for his assistance during the experiments with cubic anvil high-pressure (CA-HP) techniques.

**Table 1**: Comparison of the superconducting properties of SmFeAsO-based materials, including the critical transition temperature ($T_c$), upper critical field ($\mu_0 H_{c2}(0)$) estimated using the WHH model, Ginzburg–Landau coherence length ($\xi(0)$), and Maki parameter ($\alpha$), for different sample types (polycrystalline bulks, single crystals, thin films) and various substitution systems.

| **Sample Type** | **Compounds** | **$T_c$**<br>**(K)** | **$H_{c2}$ (0)**<br>**(T)** | **ξ (0)**<br>**(Å)** | **Maki**<br>**Paramet** | **Refs** |
| --- | --- | --- | --- | --- | --- | --- |
| Single crystal | $SmFeAsO_{0.85}$ | 55 | 380 (‖*ab*)<br>84 (‖*c*) | 20 (‖*ab*)<br>4.4 (‖*c*) | α ≈ 2.5(‖*ab*)<br>α < 1 (‖*c*) | Lee *et al.* [47] |
| | $SmFeAsO_{0.80}F_{0.20}$ | 55 | 280 (‖*ab*)<br>47(‖*c*) | 26 (‖*ab*)<br>4.4 (‖*c*) | α ≈ 3 (‖*ab*)<br>α < 1 (‖*c*) | Lee *et al.* [47] |
| Thin film | $SmFeAsO_{0.65}H_{0.35}$ | 45 | 186 (‖*ab*)<br>93 (‖*c*) | 26.5 (‖*ab*)<br>13 (‖*c*) | α ≈ 1.2(‖*ab*)<br>α < 1 (‖*c*) | Hanzawa *et al.* [48] |
| Polycrytalline | $SmFeAsO_{0.90}F_{0.10}$ | 53.8 | 201 | 12.8 | α ≈ 3 | Zhang *et al.* [57] |
| | $SmFeAsO_{0.85}F_{0.25}$ | 57.8 | 298 | 10.5 | α ≈ 4 | Singh *et al.* [5] |
| | $SmFeAsO_{0.70}F_{0.30}$ | 57.7 | 259 | 11.3 | α ≈ 3.5 | Singh *et al.* [5] |
| | $SmFeAsO_{0.85}F_{0.25}$ | 56.8 | 223 | 12. 1 | α ≈ 3 | Present study |
| | $SmFeAsO_{0.60}F_{0.40}$ | 55.5 | 125 | 17. 1 | α ≈ 2 | Present study |

**Figure 1.** **(a)** Powder X-ray diffraction (XRD) patterns of $SmFeAsO_{1-x}F_x$ samples ($0.05 \leq x \leq 0.40$) synthesized by CA-HP method along with the undoped parent sample ($x = 0$) prepared by the CSP method. The patterns are dominated by the tetragonal SmFeAs(O,F) phase with minor impurity phases identified as SmOF, SmAs, and FeAs. **(b)** Enlarged view of the (102) diffraction peak around $2\theta \approx 31°$, illustrating an overall shift towards higher diffraction angles with increasing nominal fluorine content, indicating an average lattice contraction associated with F substitution at the O site. **(c)** The Rietveld analysis for the sample $x = 0.05$ is presented, displaying the experimental and calculated diffraction patterns and their difference for the room temperature X-ray diffraction data. The compositional dependence of **(d)** lattice parameter '*a*' **(e)** lattice parameter '*c*' **(f)** unit-cell volume '*V*' extracted from Rietveld refinement is plotted as a function of nominal fluorine content $x$ ($0.05 \leq x \leq 0.40$) for the present CA-HP samples. For comparison, corresponding data for F-substituted $SmFeAsO_{1-x}F_x$ prepared by the CSP method (Singh *et al.* [5]) and for H-substituted $SmFeAsO_{1-x}H_x$ (Matsuishi *et al.* [25]) are also included in panels (d)–(f).

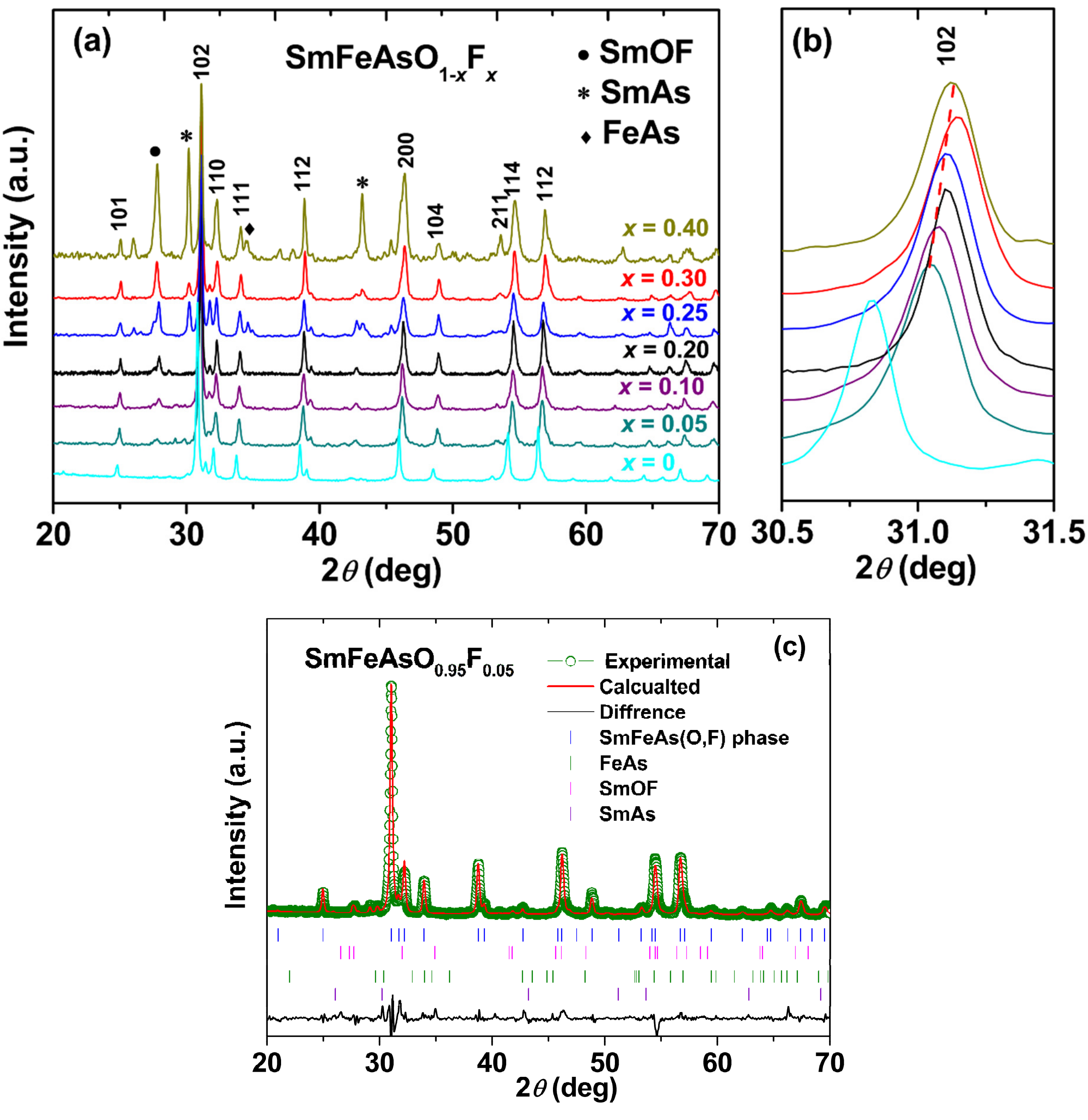

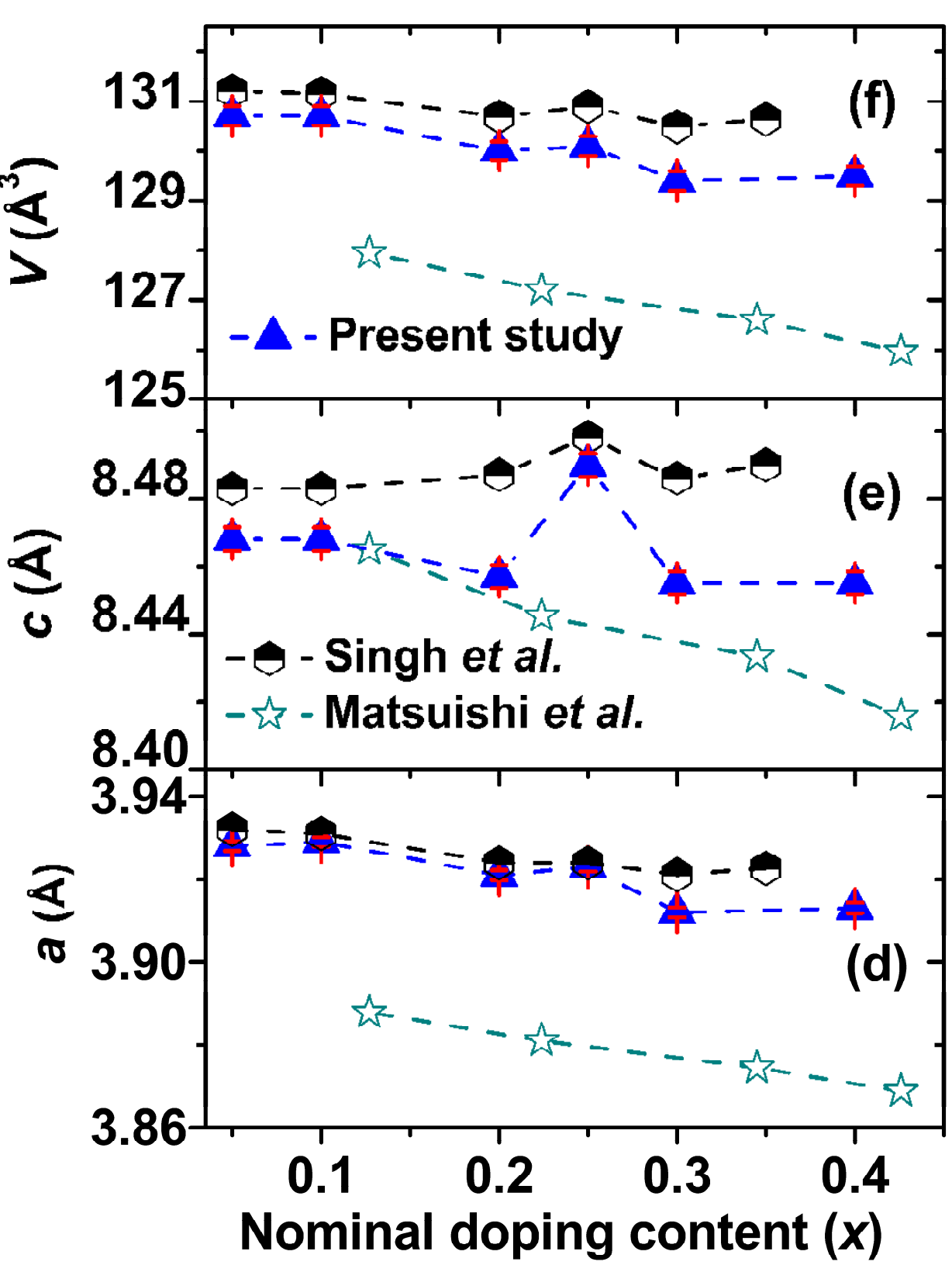

**Figure 2.** Backscattered electron (BSE) images of the representative $SmFeAsO_{1-x}F_x$ samples from the underdoped ($x$ = 0.05), optimal doped ($x$ = 0.20), and overdoped ($x$ = 0.40) regions. Images for $x$ = 0.05, 0.20, and 0.40 are shown at low magnification (10 µm; panels **(a)**,**(c)**,**(e)**) and higher magnification (200 nm; panels **(b)**,**(d)**,**(f)**). The light-grey matrix corresponds to the SmFeAs(O,F) target phase, while brighter regions are attributed to Sm-rich oxide/oxyfluoride phases ($Sm_2O_3$/SmOF). Dark regions originate from pores and/or Fe- and As-rich secondary phases such as SmAs or FeAs. Phase assignments are supported by local EDX point analysis and elemental mapping, which are provided in the Supplementary Information (Figure S2).

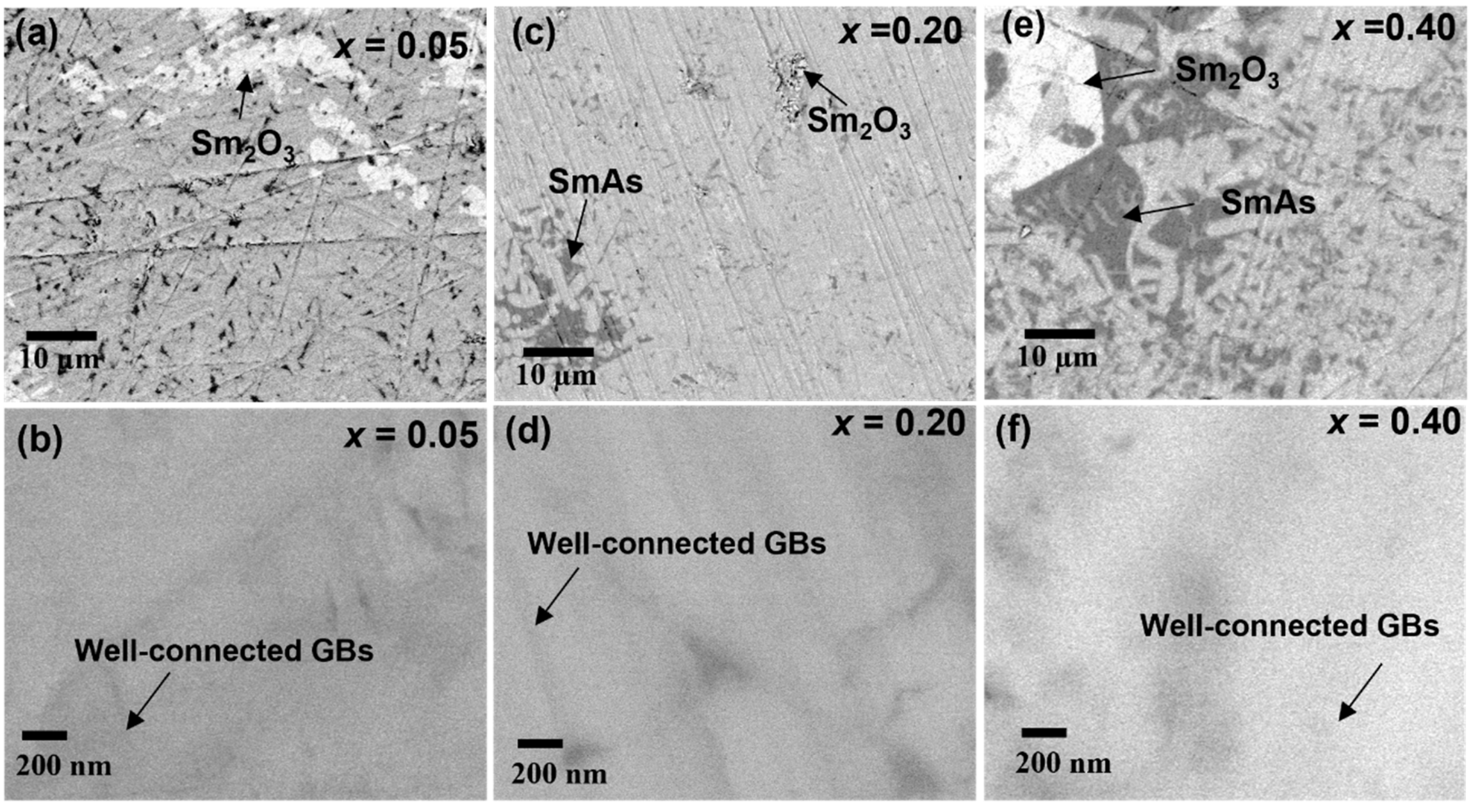

**Figure 3:** Raman scattering analysis of $SmFeAsO_{1-x}F_x$. **(a)** Representative Raman spectra for samples with the nominal fluorine concentrations ($X_F$) of 5% ($x$ = 0.05), 20% ($x$ = 0.20), and 40% ($x$ = 0.40) are presented, with each spectrum vertically displaced for better visualization. The vibrational modes corresponding to the crystal lattice are identified for the underdoped specimen. Experimental spectra were deconvoluted using Lorentzian functions, represented by green curves, while the overall Lorentzian fits to the data are depicted in red. Vertical black markers are included as visual guides. **(b)** Variation of Raman peak positions with fluorine substitution is illustrated, where the experimental values (circles) are accompanied by associated error bars. The dashed lines serve only as visual guides to indicate the general trend.

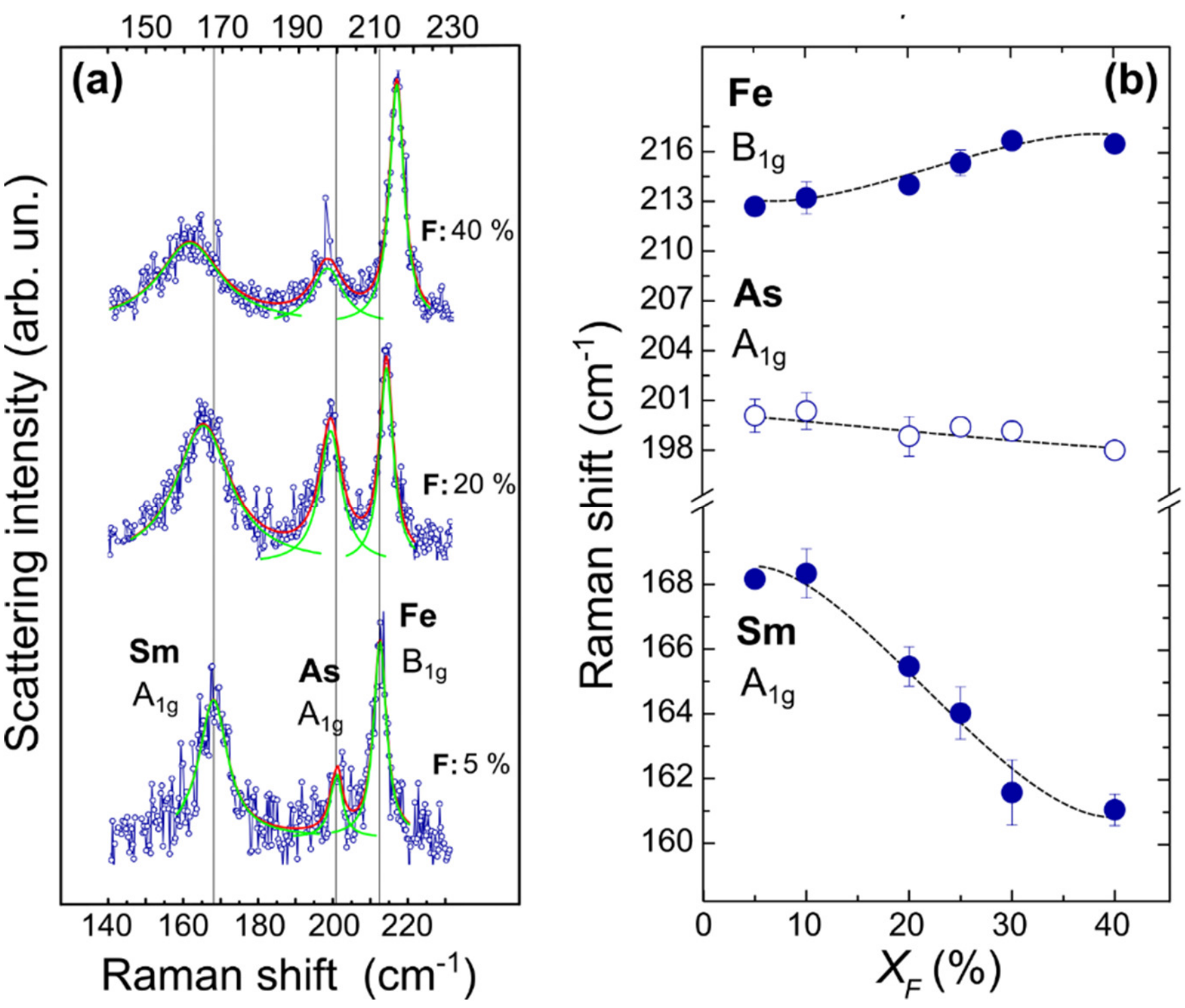

**Figure 4:** Temperature dependence of electrical resistivity $\rho(T)$ for $SmFeAsO_{1-x}F_x$ bulks. Measurements were conducted for the sample $x = 0.05$, 0.10, 0.20, 0.25, 0.30, and 0.40. Figures **(a)** and **(b)** show the $\rho(T)$ behaviour over the full temperature range up to 300 K and within the low-temperature interval of 30-60 K, respectively. The extracted transport parameters are plotted as a function of the nominal fluorine concentration ($x$) for these $SmFeAsO_{1-x}F_x$ bulks: **(c)** the onset ($T_c^{onset}$) and offset ($T_c^{offset}$) superconducting transition temperature, **(d)** the transition width ($\Delta T = T_c^{onset} - T_c^{offset}$), **(e)** the resistivity at 300 K ($\rho_{300\ K}$), and **(f)** the residual resistance ratio ($RRR = \rho_{300\ K} / \rho_{60\ K}$).

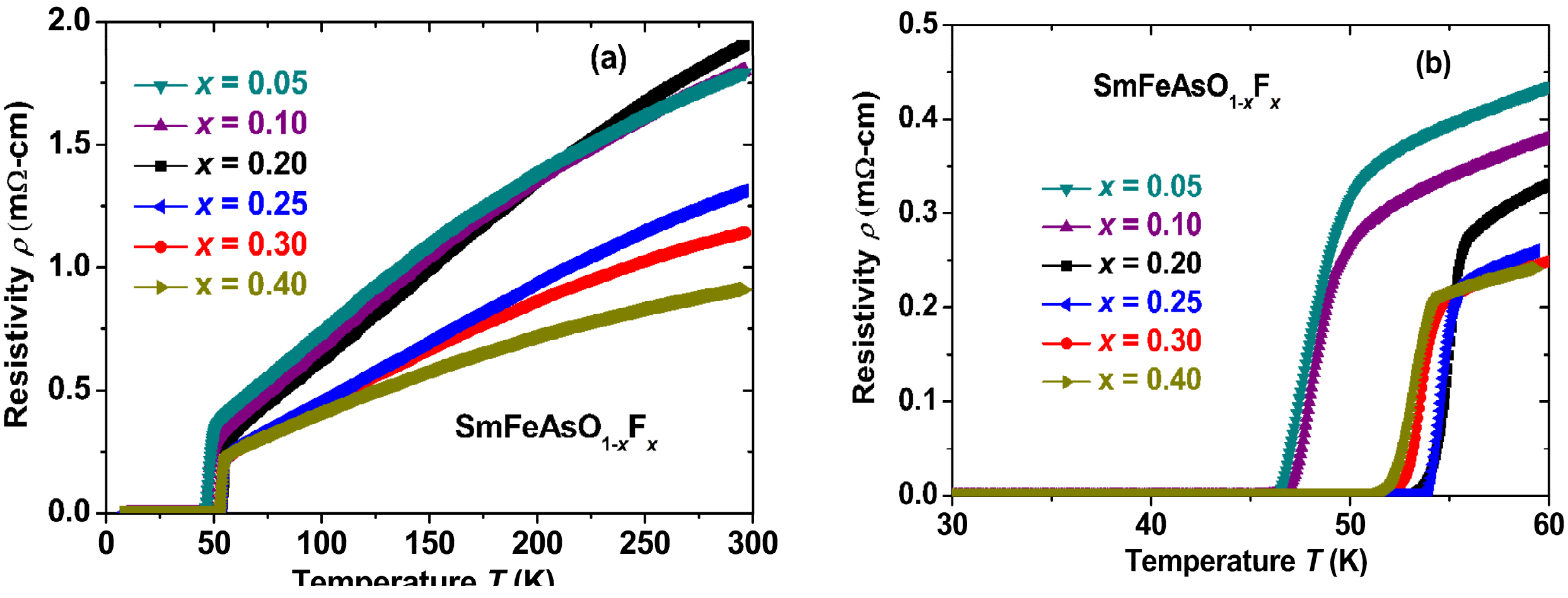

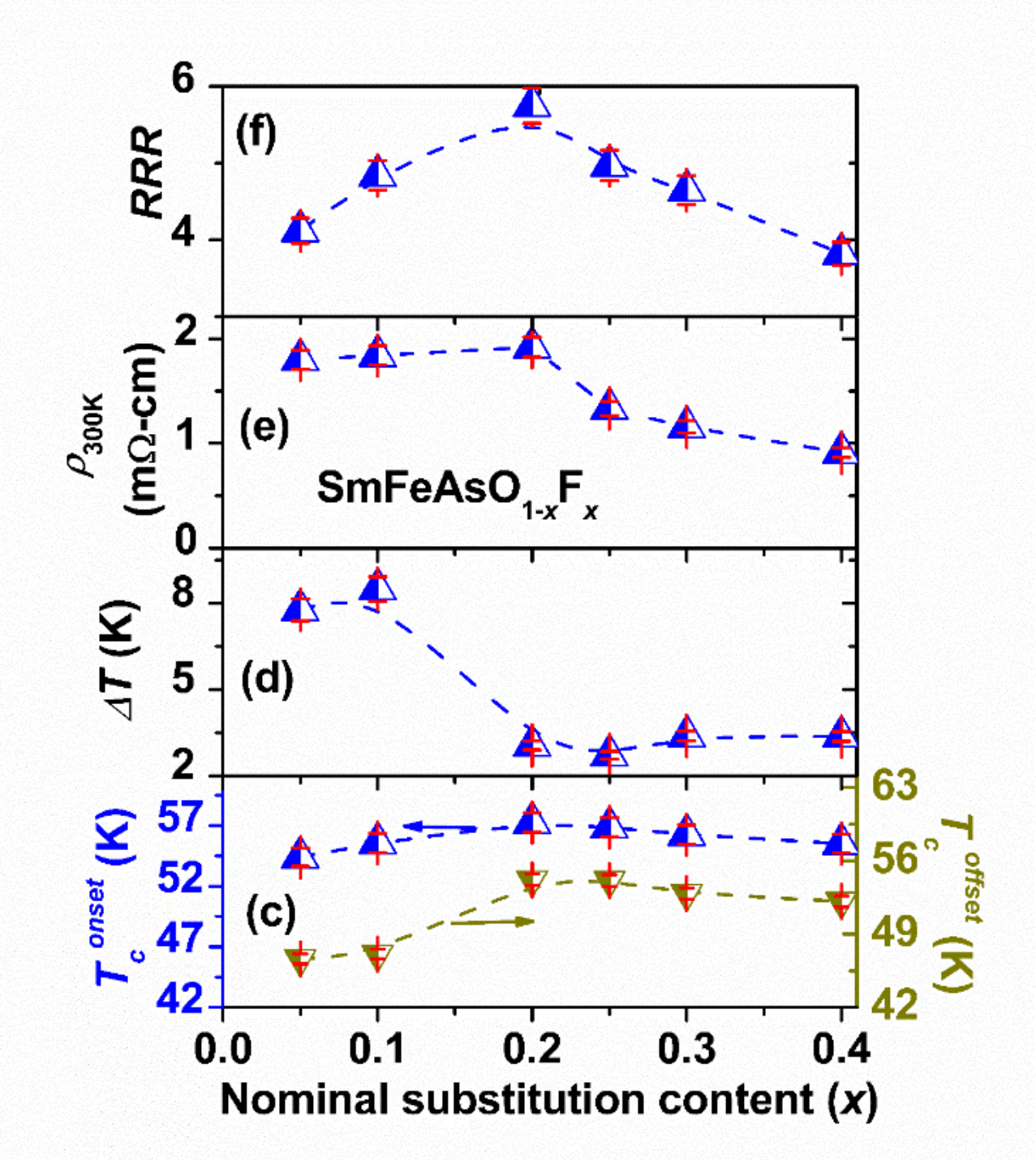

**Figure 5:** Magnetic-field-dependent resistivity of $SmFeAsO_{1-x}F_x$ bulks for **(a)** $x = 0.05$, **(b)** $x = 0.25$, and **(c)** $x = 0.40$ measured under the applied magnetic field from 0 to 9 T in the temperature range 20-70 K. The inset of each figure (a), (b) and (c) shows the corresponding *H-T* phase diagrams, where diamond symbols represent the upper critical field $H_{c2}$ (determined at 90% of the normal-state resistivity $\rho_n$) and triangle symbols denote the irreversibility field $H_{irr}$ (determined at 10% of $\rho_n$). Linear fits near $T_c$ yield slopes $dH_{c2}/dT|_{Tc} \approx$ -17.3, -5.7, -3.2 T $K^{-1}$ for $x$ = 0.05, 0.25, and 0.40, respectively.

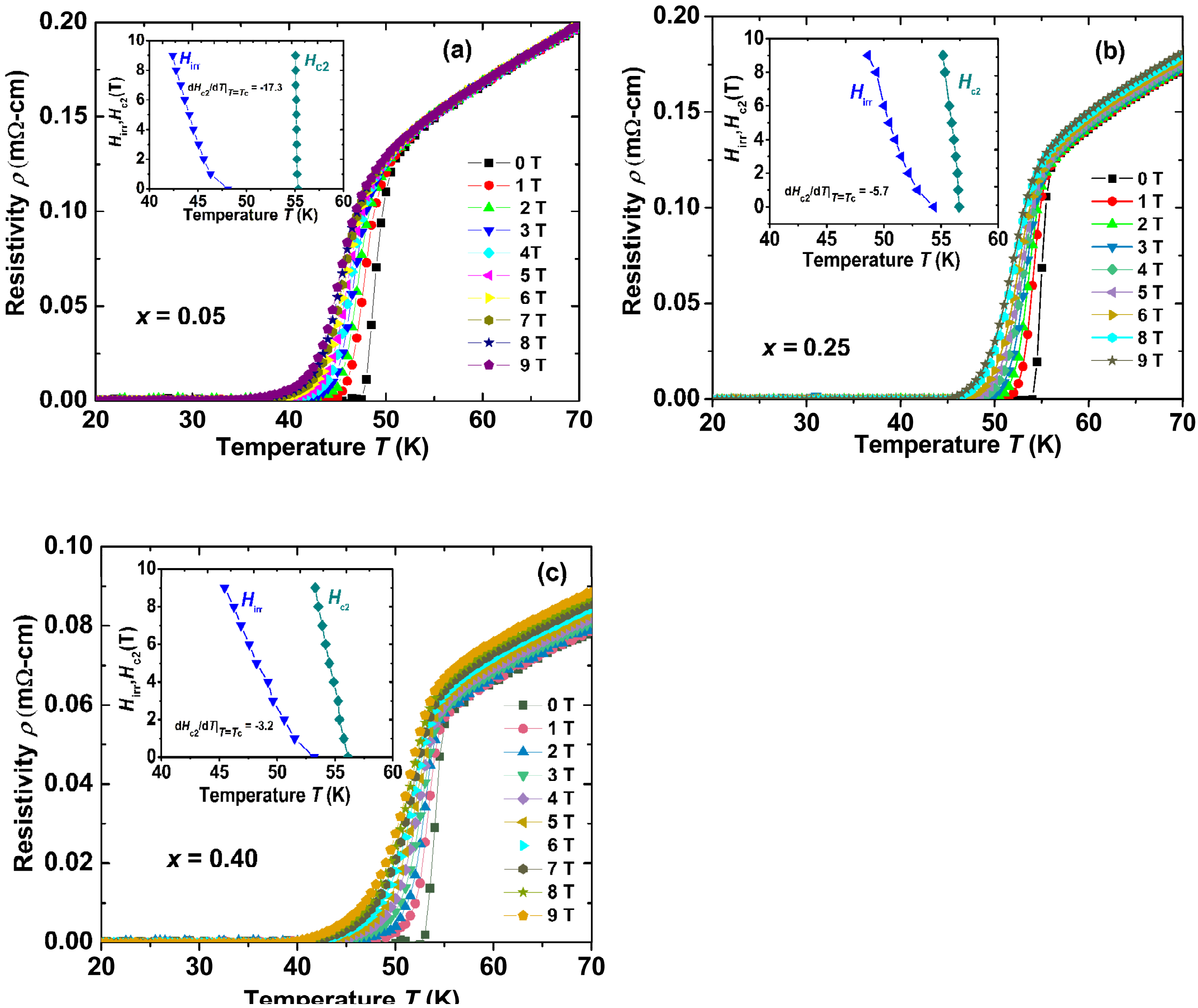

**Figure 6. (a-c)** Arrhenius plots of the normalized resistivity $\rho$ versus $1/T$ for $SmFeAsO_{1-x}F_x$ samples with $x = 0.05$, 0.25, and 0.40 measured under the magnetic field $\mu_0 H = 0$ to 9 T. Here the resistivity is normalized by the normal state resistivity $\rho_n$ just above the superconducting transition. The dashed red line in Figures (a)-(c) guides to show the linear fitting used to extract the activation energy ($U_0/k_B$). **(d)** The magnetic field dependence of the extracted activation energy $U_0/k_B$ is presented for the underdoped sample $x = 0.05$, optimal doped sample $x = 0.25$ and the overdoped sample $x = 0.40$.

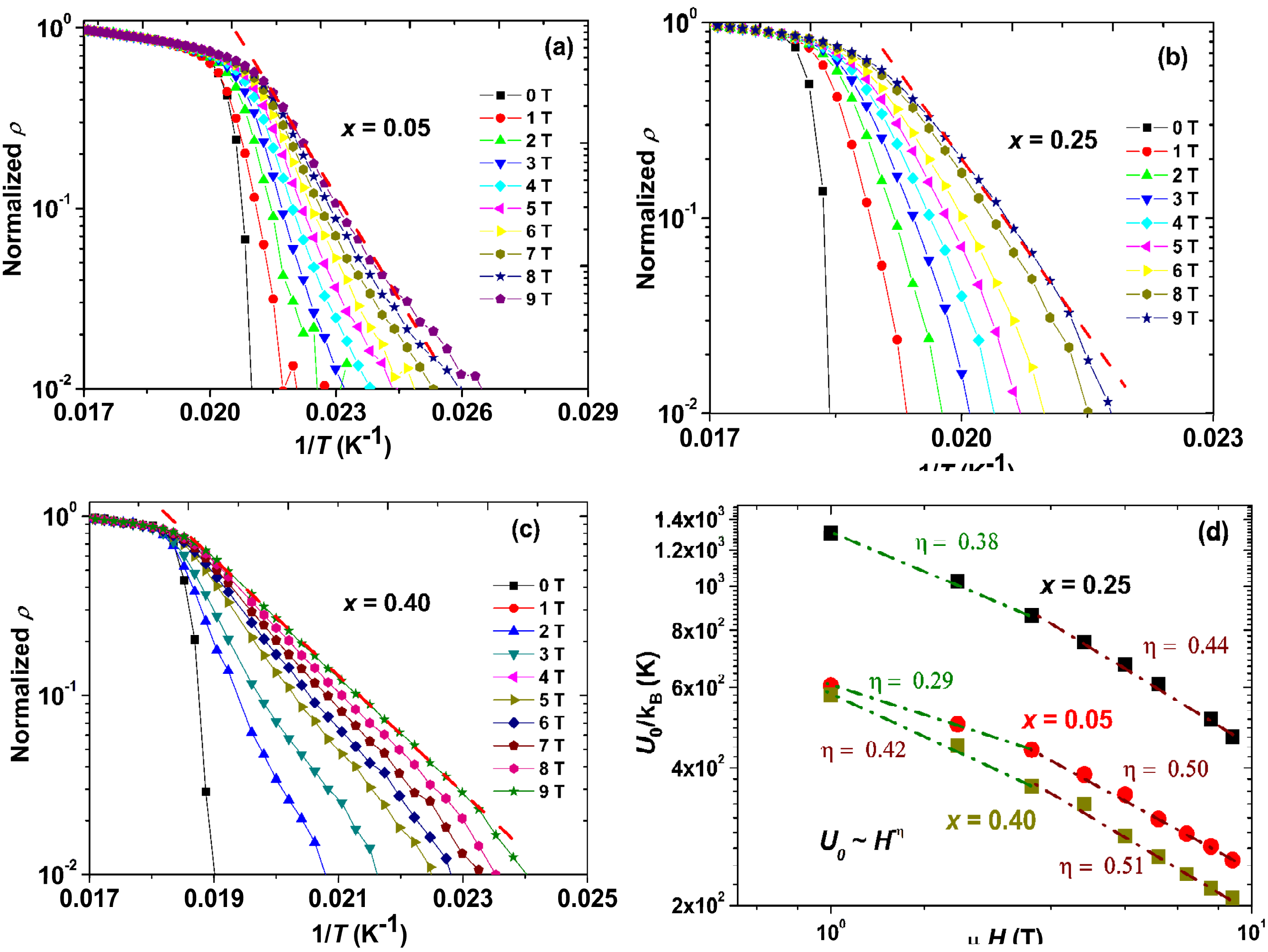

**Figure 7:** Magnetic properties of $SmFeAsO_{1-x}F_x$ bulks. **(a)** Temperature dependence of the normalized magnetic moment ($M/M_{5K}$) recorded under zero-field-cooled (ZFC) and field-cooled (FC) conditions in an applied magnetic field of 20 Oe. **(b)** Variation of the critical current density ($J_c$) at 5 K as a function of the applied magnetic fields. The inset in Figure (b) shows the field dependence of the pinning force ($F_p$) at 5 K for the samples $x = 0.05$, 0.25, and 0.40.

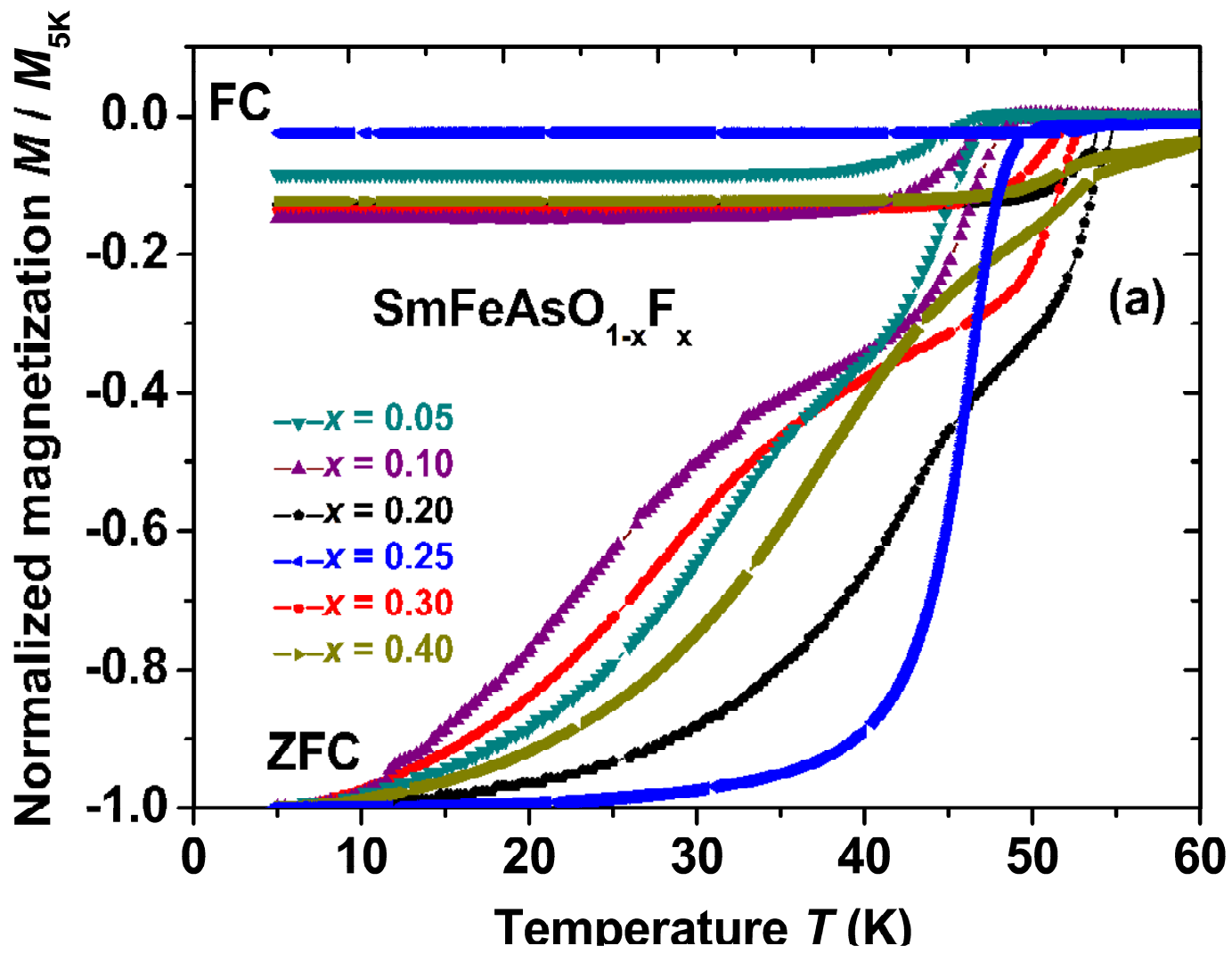

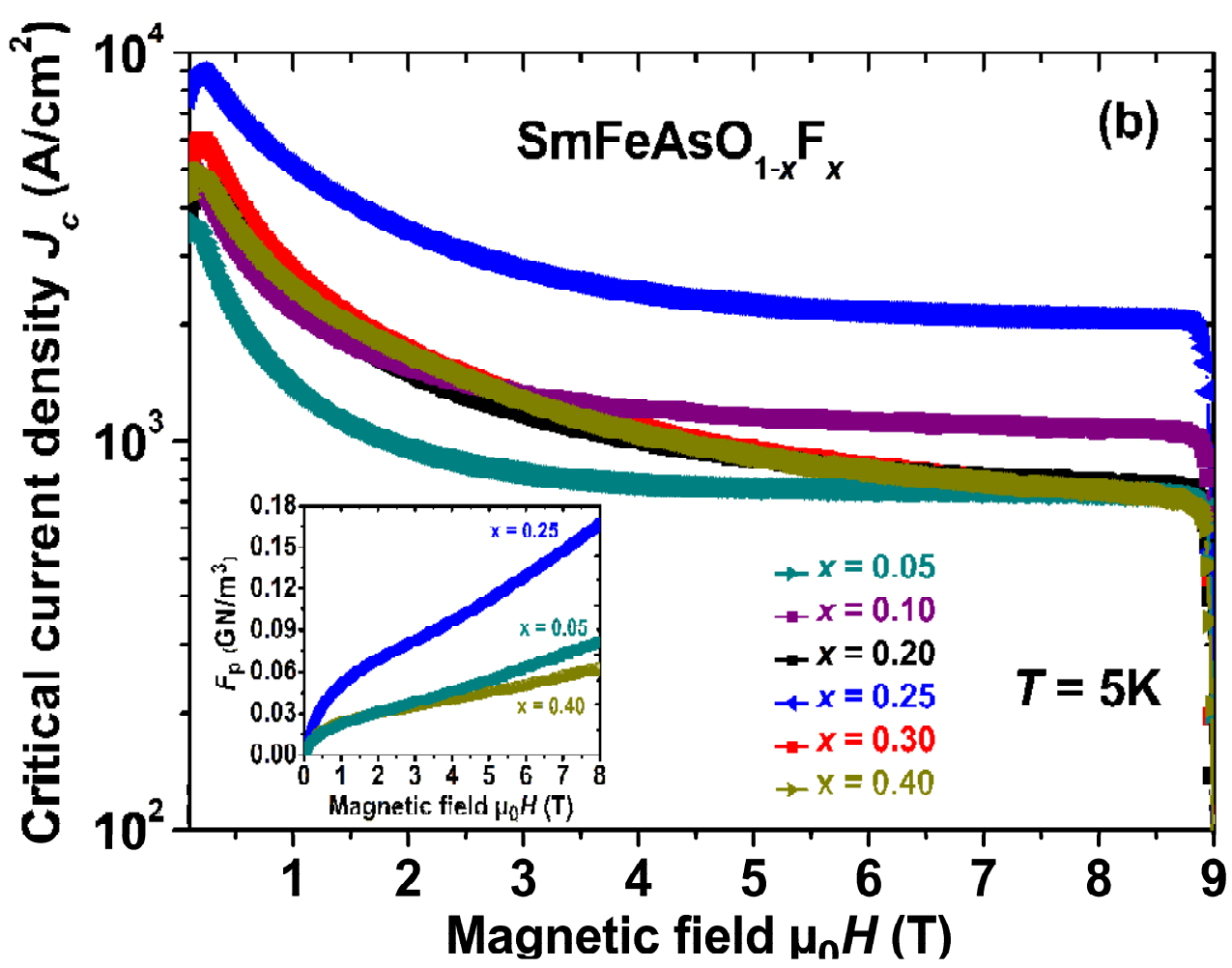

**Figure 8. (a)** Superconducting onset transition temperature $T_c$ obtained from resistivity measurements and **(b)** critical current density ($J_c$) derived from magnetization measurements for $SmFeAsO_{1-x}F_x$ samples prepared by the CA-HP process, are plotted as a function of nominal fluorine content ($x$). Three shaded colour regions depict the underdoped, optimal (Opt.) and overdoped regimes. For comparison, black hexagons denote the transport onset $T_c$ and $J_c$ (from magnetization) of $SmFeAsO_{1-x}F_x$ samples prepared by the low-temperature CSP method reported by Singh et al. [5], while triangles correspond to data for Sm1111 samples prepared by the high-temperature CSP method reported by Wang et al. [26]. Error bars for our CA-HP processed samples are included in Figures (a) and (b). In Figure (a), the onset $T_c$ values for H-substituted $SmFeAsO_{1-x}H_x$ reported by from Matsuishi et al. [25] are also shown for comparison.

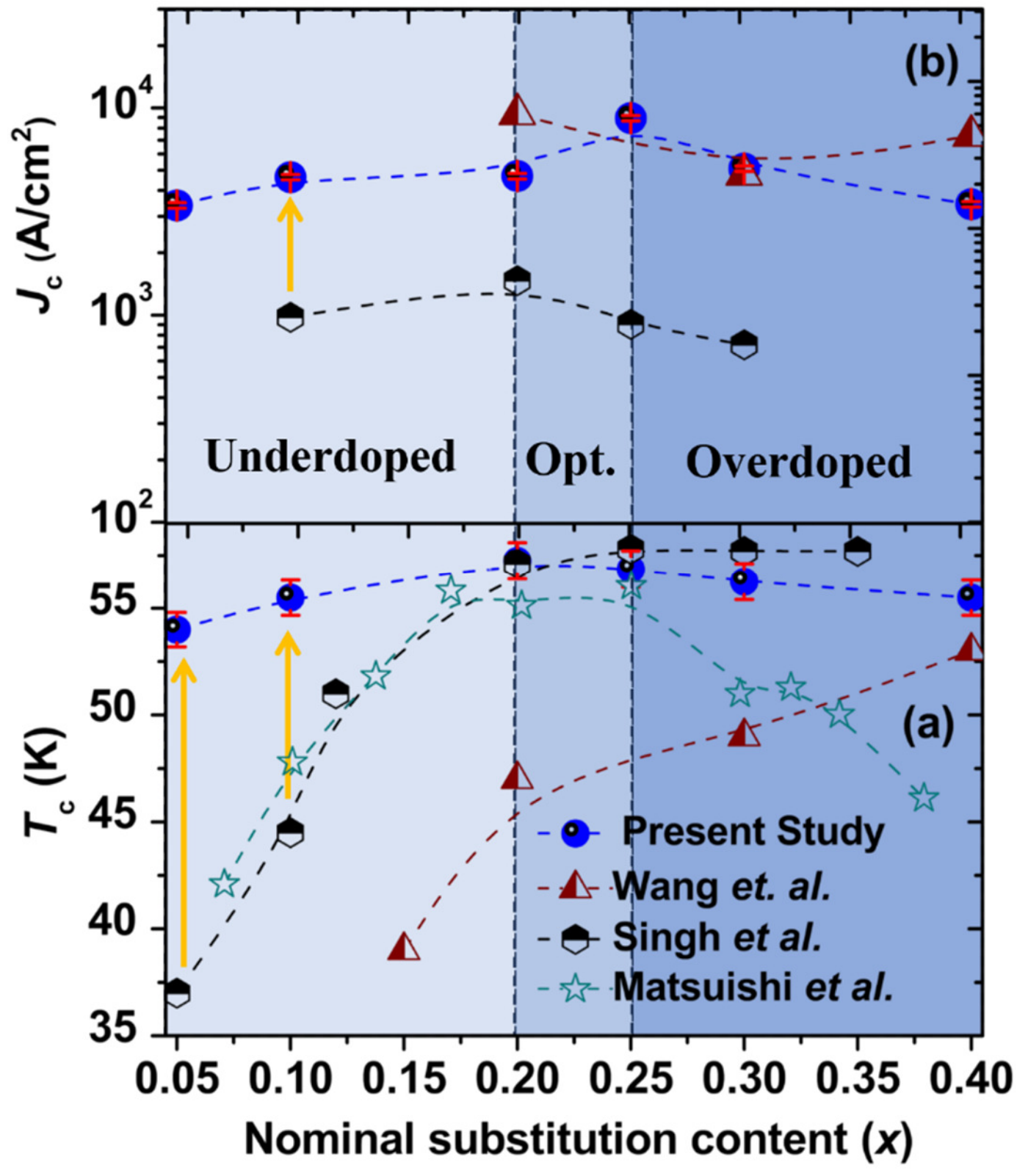